\documentclass[twocolumn,groupedaddress,aps,pre,floatfix,reprint]{revtex4-2}
\usepackage[utf8]{inputenc}
\usepackage[T1]{fontenc}

\usepackage{amsmath,amssymb} 
\usepackage{mathdots} 
\usepackage{graphicx}
\usepackage{dcolumn}
\usepackage{bm}
\usepackage{enumerate,enumitem}
\usepackage{CJK} 
\usepackage{subcaption} 
\usepackage{microtype} 
\allowdisplaybreaks
\usepackage{braket} 

\usepackage[colorlinks=true,linkcolor=blue,citecolor=blue]{hyperref}
\hypersetup{
  pdftitle={Theory of Quantum Phase Space: Foundations and Applications},
  pdfauthor={Demin Huang and Biao Wu}
}
\numberwithin{equation}{section}
\begin{document}
\title{Theory of Quantum Phase Space: Foundations and Applications}
\author{\begin{CJK}{UTF8}{gbsn} Demin Huang (黄德民) \end{CJK}}
\affiliation{International Center for Quantum Materials, School of Physics, Peking University, Beijing 100871, China}
\author{\begin{CJK}{UTF8}{gbsn} Biao Wu (吴飙) \end{CJK}}
\email{wubiao@pku.edu.cn}
\affiliation{International Center for Quantum Materials, School of Physics, Peking University, Beijing 100871, China}
\affiliation{Wilczek Quantum Center, Shanghai Institute for Advanced Studies, University of Science and Technology of China, Shanghai 201315, China}
\affiliation{Hefei National Laboratory, Hefei 230088, China}
\affiliation{Beijing Key Laboratory of Quantum Devices, Peking University, Beijing 100871, China}

\date{\today}
\begin{abstract}
This article provides a concise review of quantum phase space theory, beginning with its foundational principles and the properties 
of standard quantum quasi-probability distributions, specifically the Wigner, Husimi $Q$, and Glauber--Sudarshan $P$ functions. 
We discuss the intrinsic limitations of these distributions, such as the appearance of negative values and phase-space blurring. 
    A significant portion of this review highlights recent theoretical developments, particularly 
    the \textit{quantum Wannier basis} developed in Refs.~\cite{han2015entropy,fang2018quantum}. This approach establishes 
    a unitary mapping between the Hilbert space and a discretized phase space, yielding a genuine probability distribution 
    in phase space and thereby providing a basis-dependent entropy for pure quantum states.
    Furthermore, we examine Bourgain's nonperiodic basis as a theoretical framework to circumvent the constraints 
    imposed by the Balian--Low theorem. These developments provide practical tools for numerical studies based on the quantum 
    Wannier basis, as well as conceptual benchmarks for understanding the localization limits of orthonormal phase-space representations.
\end{abstract}
\keywords{Quasi-Probability Distribution; Wannier Basis; Bourgain's Nonperiodic Basis}
\maketitle
\tableofcontents

\vspace{0.2cm}
\noindent\rule{\linewidth}{0.4pt}
\vspace{0.2cm}

\section{Introduction}
The theory of phase space stands as one of the most profound conceptual frameworks in 
classical mechanics, offering a powerful extension beyond purely spatial descriptions of physical systems. 
In classical theory, phase space is constructed from coordinates that encompass both the positions 
and momenta of all constituent particles. The trajectories traced by these particles in phase space, 
governed by the Hamiltonian $H$, provide a lucid visualization of the system's underlying dynamics 
and its associated properties. As a result, phase-space theory has become an indispensable 
tool across diverse fields, including classical and quantum statistical physics, quantum optics, 
and quantum dynamics.

The origins of phase space can be traced back to Liouville in 1838, who demonstrated 
the conservation of volume in phase space---though without explicitly employing the term 
``phase space'' itself  \cite{liouville1838note}. Its subsequent development followed a tangled and often opaque path, 
shaped by the contributions of numerous eminent mathematicians and physicists, such as Jacobi, Poincar\'e, 
Boltzmann, and Ehrenfest \cite{PhysToday2010}. It was Gibbs who, in 1902, formally introduced 
the modern conception of phase space in his landmark work on statistical mechanics \cite{gibbs1911elementary}. 
By investigating the statistical behavior of particles within this geometric framework, 
Gibbs laid the rigorous foundations of statistical physics.

Extending the concept of phase space to the quantum domain, however, confronts a fundamental obstacle: 
the Heisenberg uncertainty principle. This principle asserts that the position and momentum of a particle 
are incompatible observables that cannot be determined simultaneously. A pioneering effort to reconcile these 
ideas was undertaken by von Neumann in 1929, who proposed a discretized phase space for representing 
wave functions \cite{neumann1929beweis,neumann2010}. He suggested partitioning phase space into ``Planck cells'' and 
sought to construct a complete orthonormal basis whose elements are each localized within a specific cell. 
Such a basis would enable a unitary mapping of wave functions onto a quantum phase space. 
By drawing on concepts from classical ensembles, von Neumann successfully derived quantum 
analogs of the ergodic theorem and the H-theorem. Nevertheless, his approach remained largely theoretical, 
as it lacked a practical method for explicitly constructing such localized bases. 
This computational hurdle, compounded by misunderstandings of his theorems \cite{Goldstein2010}, 
may have contributed to the prolonged neglect of his original proposal.

In 1932, Wigner introduced an alternative approach to quantum phase space via the 
Wigner function \cite{wigner1932quantum}. As a quasi-probability distribution, 
the Wigner function allows wave functions to be represented in phase space. Subsequent 
developments gave rise to other quasi-probability distributions, including the Husimi $Q$ function \cite{husimi1940some} 
and the Glauber--Sudarshan $P$ function \cite{glauber1963coherent,sudarshan1963equivalence}. 
These methods have found widespread application in quantum optics, nuclear and particle physics, 
condensed matter physics, and  mesoscopic systems.

Recently, von Neumann's theory of quantum phase space has been developed in 
\cite{han2015entropy,fang2018quantum}, where 
Kohn's orthogonalization method \cite{kohn1973construction} and L\"owdin's method \cite{lowdin1950non}
are combined to provide a practical construction of a complete orthonormal Wannier basis localized within each Planck cell. 
This development offers a computationally efficient realization of von Neumann's original vision. 
In contrast to the Wigner, Husimi $Q$, and Glauber--Sudarshan $P$ functions---which 
are quasi-probability distributions that may assume negative values or 
suffer from oversmoothing---the Wannier basis facilitates 
a unitary mapping that yields a non-negative, 
genuine probability distribution in quantum phase space.

The Balian--Low theorem \cite{balian1981principe} imposes a fundamental constraint on Gabor systems: 
any orthonormal basis in phase space characterized by periodic translational symmetry (i.e., a lattice structure) 
must satisfy the condition $\sigma_x \sigma_k = \infty$, where $k=p/\hbar$ denotes the wave vector.
This indicates a loss of sharp localization in either position or wave-vector space.
To circumvent this restriction, Bourgain proposed a nonperiodic basis construction that allows 
the uncertainty product to approach the Heisenberg limit arbitrarily closely, i.e., $\sigma_x \sigma_k \to \frac{1}{2}$ 
\cite{bourgain1988remark}. We provide a detailed exposition of Bourgain's nonperiodic construction.

In this review, we primarily focus on recent developments: 
the quantum Wannier basis and Bourgain's nonperiodic basis. However, we will briefly review 
the traditional frameworks, such as classical phase space and the Wigner, Husimi $Q$, and Glauber--Sudarshan $P$ 
quasi-probability distributions for the sake of comparison and self-containment.

The review is organized as follows. Section \ref{sec:2} provides a concise overview of the fundamental properties 
of classical phase space. Section \ref{sec:3} reviews the Wigner, Husimi $Q$, and Glauber--Sudarshan $P$ functions, emphasizing their 
essential characteristics and representative applications. 
Section \ref{sec:4} presents the construction of Planck-cell bases in von Neumann's quantum phase space, 
including the truncated basis and the Wannier basis obtained through Kohn and L\"owdin orthogonalization methods.
Section \ref{sec:App} illustrates their applications to the quantum kicked rotor, the two-site Bose-Hubbard model, 
quantum-state entropy, and time-frequency analysis.
Section \ref{sec:5} introduces the mathematical framework underlying Bourgain's nonperiodic basis and explores its implications 
for phase-space localization. Finally, concluding remarks are given in Section \ref{sec:6}.

\section{Classical Phase Space}
\label{sec:2}
In classical mechanics, the state of a system consisting of $N$ particles is fully characterized 
by a point in a $6N$-dimensional manifold known as phase space (or $\Gamma$-space). 
This coordinate system is spanned by the three spatial coordinates 
${q}_i$ and three conjugate momentum coordinates ${p}_i$ 
for each particle $i=1,\dots,N$.
Since classical particles possess simultaneously well-defined positions and momenta, the instantaneous 
state of the system corresponds to a unique point ${\Gamma} = (\mathbf{q}, \mathbf{p})$ in this space, where
$\mathbf{q}=(q_1,q_2,...,q_N)$ and $\mathbf{p}=(p_1,p_2,...,p_N)$.

Given a Hamiltonian $H(\mathbf{q}, \mathbf{p}, t)$, the deterministic time evolution of the system is governed by Hamilton's equations of motion:
\begin{equation}
    \frac{\mathrm{d}q_i}{\mathrm{d}t} = \frac{\partial H}{\partial p_i}, \quad \frac{\mathrm{d}p_i}{\mathrm{d}t} = -\frac{\partial H}{\partial q_i},
\end{equation}
where $i = 1, \dots, 3N$ labels the degrees of freedom. 
For a statistical ensemble of this system, we define a distribution function 
$\rho(\mathbf{q},\mathbf{p},t)$, such that 
$\rho(\mathbf{q},\mathbf{p},t)\,\mathrm{d}^{3N}q\,\mathrm{d}^{3N}p$ 
represents the probability of finding the system within the infinitesimal phase-space volume element 
$\mathrm{d}^{3N}q\,\mathrm{d}^{3N}p$ at time $t$.
The evolution of this 
density is described by the Liouville equation:
\begin{equation}
    \frac{\partial\rho}{\partial t} + \{ \rho, H \} = 0,
\end{equation}
where $\{ \cdot, \cdot \}$ denotes the Poisson bracket, defined as 
$\{ A, B \} = \sum_{i=1}^{3N} \left( \frac{\partial A}{\partial q_i} \frac{\partial B}{\partial p_i} - \frac{\partial A}{\partial p_i} \frac{\partial B}{\partial q_i} \right)$. 
This equation signifies that the local density of points in phase space remains constant as they move along their dynamical trajectories, analogous to the flow of an incompressible fluid.

One of the most significant applications of classical phase space is the qualitative description of dynamical systems 
through \textit{phase portraits}, a method pioneered by Poincar\'e \cite{poincare1886courbes}. 
A phase portrait provides a global visualization of all possible trajectories, 
effectively revealing the topological structure of the system's dynamics.

\begin{figure}[t]
    \centering
        \includegraphics[width=0.9\columnwidth]{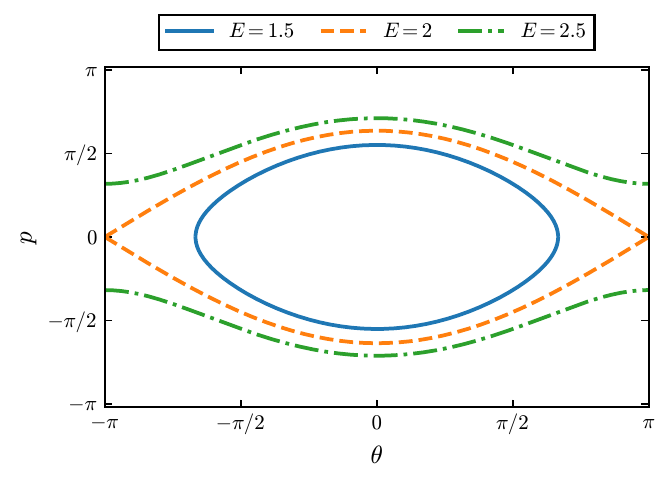}  
        \caption{Phase portrait of the integrable simple pendulum in the $(\theta,p)$ plane. The trajectories with $E=1.5$, $E=2$, and $E=2.5$ illustrate 
        the three characteristic dynamical regimes: libration below the barrier, the separatrix at the critical energy $E=2$, 
        and rotation above the barrier. The unit of energy is $mgl$.}
        \label{fig:Pendulum}
\end{figure}

Consider the example of a simple pendulum. Its motion is governed by the Hamiltonian:
\begin{equation}
    H = \frac{p^2}{2ml^2} + mgl(1 - \cos\theta),
\end{equation}
where $\theta$ denotes the angular displacement and $p$ represents the conjugate angular momentum. 
From the resulting phase space orbits as shown in Fig.~\ref{fig:Pendulum}, the periodic nature of the motion 
becomes immediately apparent.

By examining orbits at different energies, one can qualitatively distinguish between distinct dynamical 
regimes relative to the potential energy maximum, $V_{\max} = 2mgl$. 
\begin{itemize}
    \item \textbf{Libration ($E < 2mgl$):} When the total energy is below the threshold, the phase trajectories form closed loops 
    centered around the stable equilibrium ($\theta=0$). This represents periodic oscillations (libration) where the pendulum 
    remains trapped within the potential well, as exemplified by the $E = 1.5mgl$ curve in Fig.~\ref{fig:Pendulum}.
    \item \textbf{Separatrix ($E = 2mgl$):} At the critical energy level exactly equal to the potential barrier, 
    the trajectory---known as the \textit{separatrix}---connects the unstable fixed points (saddle points) at $\theta = \pm \pi$. 
    Here, the velocity of the pendulum bob vanishes as it asymptotically approaches the top position.
    \item \textbf{Rotation ($E > 2mgl$):} When the energy exceeds the barrier, the trajectories no longer close within a single cycle of $\theta \in [-\pi, \pi]$. 
    Instead, the pendulum undergoes continuous rotation with a non-zero velocity at all positions, as shown by the $E = 2.5mgl$ pattern in Fig.~\ref{fig:Pendulum}.
\end{itemize}

Although the phase trajectories for $E \geq 2mgl$ in Fig.~\ref{fig:Pendulum} 
appear to extend infinitely along the $\theta$-axis, this is an artifact of the planar representation. 
Since $\theta \equiv \theta+2\pi$, topologically identifying the boundaries at 
$\theta=-\pi$ and $\theta=\pi$ maps the phase portrait onto a cylindrical manifold.
From this perspective, the librational and rotational trajectories are periodic on the appropriate 
phase-space manifold, while the separatrix is a limiting trajectory with an infinite period.

According to the Poincar\'e-Bendixson theorem, chaos cannot occur in a continuous-time autonomous dynamical system 
with a phase-space dimension of less than three \cite{strogatz2024nonlinear}. However, chaos can 
emerge in systems that are either non-autonomous (time-dependent) or described by discrete-time maps. 
A paradigmatic example is the classical kicked rotor, which models a rigid pendulum constrained to a ring 
with periodic impulsive gravity (kicks). Its Hamiltonian is given by:
\begin{equation}
    H = \frac{p^2}{2I} + \kappa \cos x \sum_{n=-\infty}^{\infty} \delta(t - nT),
\end{equation}
where $x$ and $p$ denote the angular position and its conjugate angular momentum, $I$ is the moment of inertia, 
$\kappa$ represents the kick strength, and $T$ is the period between kicks. Between pulses ($t \neq nT$), 
there is no gravity and the rotor rotates freely; 
at each interval $T$, the kicking potential is applied instantaneously and the pendulum receives an instantaneous torque. 
Because the Hamiltonian explicitly depends on time, the system effectively possesses 
the requisite dimensionality for chaotic behavior.

\begin{figure*}[t]
    \centering
    \includegraphics[width=\textwidth]{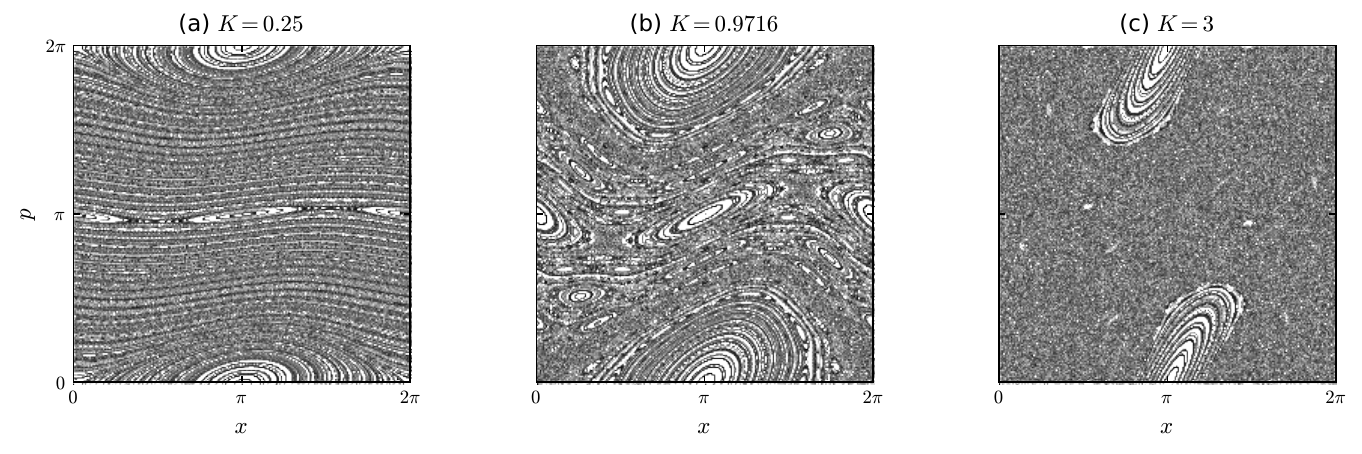}
    \caption{Poincar\'e sections of the classical kicked rotor (standard map) on the torus $(x,p)\in[0,2\pi)\times[0,2\pi)$ for 
    three representative kick strengths: (a) $K=0.25$, where the dynamics is predominantly regular and organized 
    by invariant KAM tori; (b) $K=0.9716$, corresponding to the near-critical regime in which 
    the last global invariant torus is about to break; and (c) $K=3$, where the phase space is dominated by a chaotic sea with only small regular islands surviving.}
    \label{fig:kicked_rotor_poincare}
\end{figure*}

The system can be simplified through nondimensionalization. 
By scaling variables as $t \to t/T$, $p \to pT/I$, and $H \to H T^2/I$, we obtain the dimensionless Hamiltonian:
\begin{equation}
\label{eq:hckr}
    H = \frac{p^2}{2} + K \cos x \sum_{n=-\infty}^{\infty} \delta(t - n),
\end{equation}
where $K = \kappa T/I$ is the dimensionless kick strength that fully characterizes the system's dynamical regime. 
Let $x_n$ and $p_n$ represent the state of the rotor immediately after the $n$-th kick. 
Integrating the equations of motion over one period leads to the discrete-time recursive relations:
\begin{equation}
    \begin{aligned}
        p_{n+1} &= p_n + K \sin x_n, \\
        x_{n+1} &= x_n + p_{n+1} \pmod{2\pi}.
    \end{aligned}
    \label{eq:mapckr}
\end{equation}
This set of equations is famously known as the \textit{standard map} \cite{Casati1979,smap} or the \textit{Chirikov-Taylor map}. 
The map is area-preserving, as the Jacobian determinant of the transformation is equal to unity. 
The global dynamics, ranging from near-integrable motion to global stochasticity, are determined solely by the parameter $K$.

Figure~\ref{fig:kicked_rotor_poincare}(a)--(c) shows the Poincar\'e sections of the kicked rotor for $K=0.25$, $K=0.9716$, 
and $K=3$, respectively. In these plots, periodic boundary conditions are imposed on both the $x$ and $p$ 
directions by taking $x,p \pmod{2\pi}$, so that the phase space is represented on the two-torus $\mathbb{T}^2$.

At $K=0.25$, the phase space is dominated by invariant Kolmogorov-Arnold-Moser (KAM) tori, 
indicating near-integrable behavior without global chaos. 
Figure~\ref{fig:kicked_rotor_poincare}(b) corresponds to the near-critical regime $K_c \approx 0.9716$. 
This value is historically significant as the golden-mean threshold, representing the largest kick strength at which the last 
global invariant torus persists; 
beyond this value, global invariant barriers to momentum transport are destroyed. At $K=3$, the phase space is 
dominated by a chaotic sea, where most trajectories are stochastic and only small regular islands remain. 
This evolution clearly demonstrates the transition from integrability to global chaos in a Hamiltonian system. 
For a more comprehensive discussion of classical and quantum kicked rotors, we refer the reader to \cite{santhanam2022quantum} 
and Strogatz's seminal text \cite{strogatz2024nonlinear}.

However, the deterministic trajectories of classical phase space are fundamentally incompatible with quantum mechanics due to the 
Heisenberg uncertainty principle. Since a quantum particle cannot possess a simultaneously definite position and momentum, 
the concept of a ``point'' in phase space must be replaced by a distribution. Various frameworks have been developed to represent 
the state of a quantum system in this quasi-classical language, most notably the Wigner, Husimi $Q$, and Glauber--Sudarshan $P$ functions, 
as well as the recently developed Wannier basis. In the following sections, we first examine the properties and limitations of the three quantum quasi-probability distributions.

As the extension from one-dimensional to $d$-dimensional systems is mathematically straightforward, 
this review primarily focuses on one-dimensional cases.

\section{Quantum Quasi-probability Distributions}
\label{sec:3}
\subsection{The Wigner Function}

\subsubsection{Definition and Formalism}
The Wigner function (often referred to as the $W$-function) is a seminal quasi-probability distribution introduced 
by Wigner in 1932 \cite{wigner1932quantum}. It was designed to bridge the gap between 
the Hilbert space formalism of quantum mechanics and the classical phase-space description, 
effectively mapping quantum states onto classical phase space. 
For a pure state described by the wave function $\psi(x)$ in the coordinate representation, 
the Wigner function $W(x,p)$ is defined as:
\begin{equation}
    W(x,p) := \frac{1}{\pi\hbar} \int_{-\infty}^{+\infty} \mathrm{d}y \, \psi^{*}(x+y) \psi(x-y) e^{\frac{2ipy}{\hbar}},
\end{equation}
where $x$ and $p$ correspond to the eigenvalues of the position operator $\hat{x}$ and momentum operator $\hat{p}$, respectively. 

The formalism naturally extends to statistical ensembles described by a density operator $\hat{\rho} = \sum_{i} \lambda_i \ket{\psi_i}\bra{\psi_i}$, 
where $\lambda_i$ represents the probability of the system being in state $\ket{\psi_i}$. In this general case, 
the Wigner function is expressed as the partial Fourier transform of the off-diagonal elements of the density matrix:
\begin{equation}
    W(x,p) = \frac{1}{\pi\hbar} \int_{-\infty}^{+\infty} \mathrm{d}y \, \bra{x-y} \hat{\rho} \ket{x+y} e^{\frac{2ipy}{\hbar}}.
\end{equation}
Here, $\ket{x\pm y}$ are the eigenstates of the position operator. For systems in higher dimensions, such as a particle 
in three-dimensional space, the Wigner function is defined over a six-dimensional phase space as:
\begin{equation}
    W(\mathbf{x},\mathbf{p}) = \frac{1}{(\pi\hbar)^3} \int_{\mathbb{R}^3} \mathrm{d}^{3}y \, \psi^{*}(\mathbf{x}+\mathbf{y}) \psi(\mathbf{x}-\mathbf{y}) e^{\frac{2i\mathbf{p}\cdot\mathbf{y}}{\hbar}}.
\end{equation}

\subsubsection{Elementary Properties of the Wigner Function}
The Wigner function possesses several fundamental properties that underscore its role as a phase-space representation of quantum states.

\begin{enumerate}[label=(\roman*)]
    \item \textbf{Symmetry between $x$ and $p$:} 
    \label{pro:symmetry}
    The Wigner function exhibits a high degree of symmetry between the position and momentum representations. It can be equivalently expressed using the momentum-space wave function $\varphi(p) = \langle p | \psi \rangle$:
    \begin{equation}
        W(x,p) = \frac{1}{\pi\hbar} \int_{-\infty}^{+\infty} \mathrm{d}q \, \varphi^{*}(p+q) \varphi(p-q) e^{-\frac{2iqx}{\hbar}}.
    \end{equation}
    The symmetry highlights that the $W$-function treats $x$ and $p$ on an essentially equal footing.

    \item \textbf{Reality:} 
    $W(x,p)$ is strictly real-valued for any state. By taking the complex conjugate:
    \begin{equation}
        W^*(x,p) = \frac{1}{\pi\hbar} \int_{-\infty}^{+\infty} \mathrm{d}y \, \psi(x+y) \psi^{*}(x-y) e^{-\frac{2ipy}{\hbar}}.
    \end{equation}
    Performing a variable substitution $y \to -y$, the integral returns to its original form, proving $W^*(x,p) = W(x,p)$. 
    This reality is a prerequisite for its interpretation as a quasi-probability density.
    
     \item \textbf{Marginal Distributions and Normalization:}
    A key property of $W(x,p)$ is that its marginals yield the correct quantum probability densities in position and momentum space:
    \begin{equation}
        \int W(x,p) \mathrm{d}p = |\psi(x)|^2, \quad \int W(x,p) \mathrm{d}x = |\varphi(p)|^2.
    \end{equation}
    Consequently, the Wigner function is normalized: $\iint W(x,p) \mathrm{d}x \mathrm{d}p = 1$. 
    These properties allow $W(x,p)$ to behave like a standard probability distribution when integrated over one entire dimension.

    \item \textbf{Boundedness:} 
    The Wigner function is pointwise bounded by the value inverse to the half-Planck constant:
    \begin{equation}
        -\frac{1}{\pi\hbar} \leq W(x,p) \leq \frac{1}{\pi\hbar}.
    \end{equation}
    This can be proved using the Cauchy-Schwarz inequality.  
    \item \textbf{Existence of Negative Values:} 
    \label{pro:negative}
    Unlike classical probability densities, the Wigner function can take negative values, which serves as a signature of 
    non-classicality and quantum interference. Consider the first excited state ($n=1$) of a quantum harmonic oscillator (using dimensionless oscillator units, $m=\omega=\hbar=1$):
    \begin{equation}
        \psi(x) = \sqrt{\frac{2}{\sqrt{\pi}}} x e^{-x^2/2}.
    \end{equation}
    The corresponding Wigner function is:
    \begin{equation}
        W(x,p) = \frac{1}{\pi} (2x^2 + 2p^2 - 1) e^{-(x^2+p^2)}.
    \end{equation}
    At the origin of the phase space, $W(0,0) = -1/\pi < 0$. Such negative regions highlight the ``quasi'' nature of the distribution.

    \item \textbf{Nonlinearity and Interference:}
    The Wigner function is a quadratic functional of the state. 
    For a superposition state $\psi(x) = c_1 \psi_1(x) + c_2 \psi_2(x)$, the resulting Wigner function is:
    \begin{equation}
        W(x,p) = |c_1|^2 W_1(x,p) + |c_2|^2 W_2(x,p) + I_{12}(x,p),
    \end{equation}
    where the interference term (or cross term) is given by
    \begin{align}
        &~~~~I_{12}(x,p)  \nonumber\\
        &=\frac{2}{\pi\hbar} \text{Re} \left[ \int_{-\infty}^{+\infty} \mathrm{d}y \, c_1^* c_2 \psi_1^*(x+y) \psi_2(x-y) e^{\frac{2ipy}{\hbar}} \right].
    \end{align}
    The existence of $I_{12}$ is responsible for the complex oscillating structures often observed in the phase space of cat states.

    \item \textbf{Operator Mapping and Reversibility:}
    While the Wigner function is insensitive to the global phase of a wave function $\psi(x)$, the transformation is fully reversible at the level of operators. 
    Any Hilbert space operator $\hat{A}$ can be mapped to a phase-space function $W_A(x,p)$ (its \textit{Wigner-Weyl symbol}):
    \begin{equation}
        W_A(x,p) = \frac{1}{\pi\hbar} \int_{-\infty}^{+\infty} \mathrm{d}y \, \bra{x-y} \hat{A} \ket{x+y} e^{\frac{2ipy}{\hbar}}.
    \end{equation}
    The inverse mapping, known as the \textit{Weyl transform}, recovers the operator from its symbol:
    \begin{equation}
        \hat{A} =  \iint \mathrm{d}x \mathrm{d}p \, W_A(x,p) \hat{\Delta}(x,p),
    \end{equation}
    where $\hat{\Delta}(x,p)$ is the \textit{phase-space kernel operator}, also referred to as the Stratonovich--Weyl kernel. 
    To provide a complete description of this isomorphism, the kernel operator is explicitly defined as the Fourier transform of the displacement operator:
    \begin{equation}
        \hat{\Delta}(x,p) = \frac{1}{2\pi\hbar} \iint \mathrm{d}u \mathrm{d}v \, e^{\frac{i}{\hbar}(u(\hat{p}-p) - v(\hat{x}-x))}.
    \end{equation}
    Equivalently, it can be expressed in terms of the parity operator $\hat{\Pi}$ (which performs a reflection about the origin) shifted to the point $(x,p)$ in phase space:
    \begin{equation}
        \hat{\Delta}(x,p) = 2 \hat{D}(x,p) \hat{\Pi} \hat{D}^\dagger(x,p),
    \end{equation}
    where $\hat{D}(x,p)=\exp\!\left[\frac{i}{\hbar}(p\hat{x}-x\hat{p})\right]$ is the phase-space displacement operator.

    This establishes a rigorous isomorphism between the operator algebra in Hilbert space and the function algebra in phase space, 
    often called the \textit{Moyal star-product algebra}. Under this mapping, the trace of the product of two operators $\hat{A}$ and $\hat{B}$ satisfies the overlap relation (Moyal's formula):
    \begin{equation}
         \mathrm{Tr}(\hat{A}\hat{B}) = 2\pi\hbar \iint \mathrm{d}x \mathrm{d}p \, W_A(x,p) W_B(x,p).
    \end{equation}
    This ensures that expectation values of observables $\langle \hat{A} \rangle =  \mathrm{Tr}(\hat{\rho}\hat{A})$ can be calculated 
    as classical-like phase-space averages $\iint W_\rho W_A \mathrm{d}x \mathrm{d}p$, further justifying the utility of the Wigner representation \cite{moyal1949quantum,case2008wigner}.

    \item \textbf{Time Evolution and the Moyal Equation:}
    The dynamics of the Wigner function are governed by the Moyal equation:
    \begin{equation}
        \frac{\partial W}{\partial t} = \{H, W\}_\star = \frac{1}{i\hbar} (H \star W - W \star H),
        \label{equ:Wigner_time}
    \end{equation}
    where $\star$ denotes the Groenewold--Moyal star product. For one-dimensional systems:
    \begin{equation}
        (A \star B)(x,p) = A(x,p) \exp \left[ \frac{i\hbar}{2} \left( \overleftarrow{\partial}_x \overrightarrow{\partial}_p - \overleftarrow{\partial}_p \overrightarrow{\partial}_x \right) \right] B(x,p),
    \end{equation}
    where the arrows indicate the direction in which the derivatives act.
    For  free particles or harmonic oscillators, 
    the Moyal equation simplifies to the classical-like form \cite{case2008wigner}:
    \begin{equation}
        \frac{\partial W}{\partial t} = -\frac{p}{m} \frac{\partial W}{\partial x} + \frac{\partial V(x)}{\partial x} \frac{\partial W}{\partial p}.
    \end{equation}
    In general, in the classical limit $\hbar \to 0$, the Moyal bracket $\{ \cdot, \cdot \}_\star$ reduces 
    to the classical Poisson bracket, and Eq.~\eqref{equ:Wigner_time} converges to the Liouville equation. 
    When the density in phase space is a delta function, the Liouville equation is reduced to Newton's equations of motion. 
    This indicates that when the Wigner function is very narrow in phase space, its time evolution is expected to
     follow closely the classical motion. 
\end{enumerate}
These are the main properties of the Wigner function.
More details about the Wigner function can be found in \textit{Handbook of function and generalized function transformations} \cite{zayed1996handbook}.

\subsubsection{Example: The Quantum Harmonic Oscillator}
To illustrate  the Wigner representation, we consider  the energy eigenstates of a quantum harmonic oscillator, which
are also called Fock states. The wave function for the $n$-th eigenstate is given by:
\begin{equation}
    \psi_n(x) = \left( \frac{m\omega}{\pi\hbar} \right)^{1/4} \frac{1}{\sqrt{2^n n!}} H_n\left( \sqrt{\frac{m\omega}{\hbar}} x \right) e^{-\frac{m\omega}{2\hbar} x^2},
\end{equation}
where $H_n(\xi)$ denotes the physicists' Hermite polynomial of order $n$, defined by $H_n(\xi) = (-1)^n e^{\xi^2} \frac{\mathrm{d}^n}{\mathrm{d}\xi^n} e^{-\xi^2}$.

The corresponding Wigner function for the $n$-th state is expressed in terms of Laguerre polynomials \cite{davies1975wigner}:
\begin{equation}
\begin{split}
    W_n(x,p) &= \frac{(-1)^n}{\pi\hbar} L_n\left( \frac{4H_{\mathrm{cl}}}{\hbar\omega} \right) \exp\left( -\frac{2H_{\mathrm{cl}}}{\hbar\omega} \right), \\
    \label{equ:W_harmonic}
\end{split}
\end{equation}
where $H_{\mathrm{cl}} = \frac{p^2}{2m} + \frac{1}{2}m\omega^2 x^2$ is the classical Hamiltonian, and $L_n(\zeta)$ is the $n$-th order 
Laguerre polynomial defined as $L_n(\zeta) = \frac{e^\zeta}{n!} \frac{\mathrm{d}^n}{\mathrm{d}\zeta^n} (\zeta^n e^{-\zeta})$.

\begin{figure}[htbp]
    \centering
    \includegraphics[width=0.85\columnwidth]{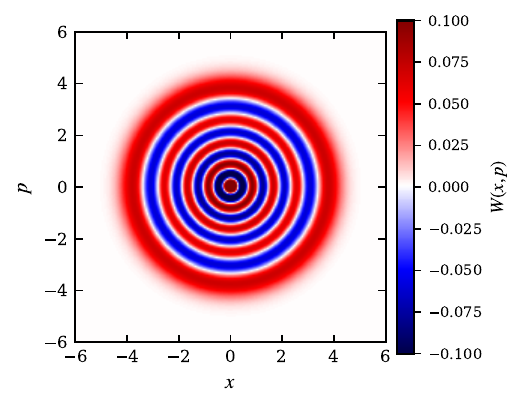}
    \caption{The Wigner function of the $8$-th energy eigenstate of a harmonic oscillator ($n=8$). 
    The presence of significant negative regions near the origin is a hallmark of the state's non-classicality.}
    \label{fig:Wigner_harmonic}
\end{figure}

As illustrated in Fig.~\ref{fig:Wigner_harmonic} for the $n=8$ Fock state, 
the Wigner function exhibits a series of concentric circular ripples in phase space. 
In this representation, the transition from red to blue signifies the shift from positive to negative quasi-probability density, 
respectively. These alternating regions of positivity and negativity encapsulate a fundamental aspect of the Wigner representation: 
they are a direct manifestation of the strong quantum interference intrinsic to Fock states. In the classical limit of large $n$, these oscillations 
become increasingly rapid, and the distribution concentrates around the classical trajectory defined by $H_{\mathrm{cl}} = (n+\frac{1}{2})\hbar\omega$.  

This visualization in Fig.~\ref{fig:Wigner_harmonic} 
vividly demonstrates several fundamental properties of the Wigner representation:
\begin{enumerate}[label=(\roman*)]

\item The prominent blue regions exemplify property \ref{pro:negative}. According to Hudson's theorem \cite{hudson1974wigner}, 
the only pure states with a non-negative Wigner function are Gaussian states, which include the ground state 
and coherent states \cite{hillery1984distribution}. Any other states necessarily exhibit negative regions, a hallmark of its non-classical nature.

\item The structural topology of the distribution is governed by the $L_n$ Laguerre polynomial in Eq.~\eqref{equ:W_harmonic}. 
Beyond the central peak, the number of surrounding concentric oscillatory rings precisely matches the excitation order $n$. 
This nodal structure is a direct phase-space manifestation of the spatial nodes in the wave function.

\item We observe that the Wigner function reaches its maximum absolute value at the origin $(0,0)$, while a series 
of subsidiary annular peaks extend outward. The outermost ring, where the distribution is most concentrated, 
qualitatively corresponds to the classical trajectory--a circle in the scaled phase space representing constant energy $E_n = (n+1/2)\hbar\omega$. 
In contrast, the intense oscillatory behavior and the central peak near the origin have no classical counterparts; they represent purely quantum interference effects, 
illustrating the intricate structures inherent in quasi-probability distributions.
\end{enumerate}

\subsection{The Husimi \texorpdfstring{$Q$}{Q} Function}
\subsubsection{Definition and Coherent State Formalism}
The Husimi $Q$ function (hereafter referred to as the $Q$ function) was introduced 
by Husimi in 1940 \cite{husimi1940some}. It represents another important quasi-probability distribution 
utilized in many fields, including quantum optics and quantum statistical physics. 
A defining characteristic of the $Q$ function is its non-negativity across the entire phase space. However, it is not a 
true probability distribution in the classical sense because its marginal distributions do not recover 
the exact quantum probability densities $|\psi(x)|^2$ or $|\varphi(p)|^2$ due to the intrinsic ``blurring'' caused by the uncertainty principle.

The construction of the $Q$ function relies on the formalism of coherent states, which were pioneered 
by Schr\"odinger \cite{schrodinger1926stetige} as quantum states that closely mimic classical dynamics.
Coherent states saturate the Heisenberg uncertainty relation,
$\Delta x\Delta p=\hbar/2$,
and are therefore often regarded as ``the most classical'' quantum states.

A coherent state $\ket{\alpha}$ is defined as the eigenstate of the annihilation operator $\hat{a}$ \cite{klauder1960action}:
\begin{equation}
    \hat{a}\ket{\alpha} = \alpha\ket{\alpha},
\end{equation}
where $\alpha$ is a complex number and the annihilation operator is given by 
$\hat{a} = \sqrt{\frac{m\omega}{2\hbar}} \left( \hat{x} + \frac{i}{m\omega}\hat{p} \right)$. In the position representation, 
the wave function of $\ket{\alpha}$ is a Gaussian wave packet:
\begin{equation}
    \braket{x|\alpha} = \left( \frac{m\omega}{\pi\hbar} \right)^{1/4} \exp \left[ -\frac{m\omega}{2\hbar}(x-x_c)^2 + \frac{i}{\hbar} p_c \left(x - \frac{x_c}{2}\right) \right],
\end{equation}
where the phase-space coordinates $(x_c, p_c)$ are related to  $\alpha$ via:
\begin{equation}
    x_c = \sqrt{\frac{2\hbar}{m\omega}} \text{Re}(\alpha), \quad p_c = \sqrt{2\hbar m\omega} \text{Im}(\alpha).
\end{equation}
Physically, $\ket{\alpha}$ can be viewed as the ground state of a harmonic oscillator translated to the point $(x_c, p_c)$ in phase space.

The coherent state can be expanded in the Fock basis $\{\ket{n}\}$ as:
\begin{equation}
    \ket{\alpha} = e^{-\frac{1}{2}|\alpha|^2} \sum_{n=0}^{\infty} \frac{\alpha^n}{\sqrt{n!}} \ket{n},
    \label{equ:Husimi_superpostion}
\end{equation}
where the factor $e^{-|\alpha|^2/2}$ ensures normalization, and $\ket{n}$ is the number state satisfying
$\hat{a}^\dagger\hat{a}\ket{n}=n\ket{n}$.

Due to the non-vanishing overlap of Gaussian wave packets across the entire phase space, any two coherent states with 
distinct eigenvalues are not strictly orthogonal. The overlap between two coherent states $\ket{\alpha}$ and $\ket{\beta}$ is given by:
\begin{equation}
        \Braket{\alpha|\beta} = \exp \left( -\frac{1}{2}|\alpha|^2 - \frac{1}{2}|\beta|^2 + \alpha^*\beta \right), 
 \end{equation}
 which leads to 
 \begin{equation}
        |\Braket{\alpha|\beta}|^2 = \exp(-|\alpha-\beta|^2).
        \label{co_orth}
\end{equation}
This result, derived from Gaussian integration, shows that the overlap probability $|\Braket{\alpha|\beta}|^2$ decays exponentially with the squared distance between the points in phase space.

A fundamental property of the coherent state manifold is its \textit{overcompleteness}, implying that any quantum state 
can be expanded in terms of coherent states, albeit not uniquely. The set satisfies the \textit{resolution of unity} (completeness relation):
\begin{equation}
    \frac{1}{\pi} \int \mathrm{d}^2\alpha \, \ket{\alpha}\bra{\alpha} = \hat{1},
    \label{equ:Husimi_normalization}
\end{equation}
which can be proved by using Fock expansion Eq.~\eqref{equ:Husimi_superpostion} and  
polar coordinates $\alpha = r e^{i\theta}$. 

With these properties established, we formally define the Husimi $Q$ function for a general density operator $\hat{\rho}$ as:
\begin{equation}
    Q(\alpha) = \frac{1}{\pi} \Braket{\alpha | \hat{\rho} | \alpha}.
\end{equation}
For a pure state $\hat{\rho} = \ket{\psi}\bra{\psi}$, this reduces to the scaled squared overlap with the coherent state:
\begin{equation}
    Q(\alpha) = \frac{1}{\pi} |\Braket{\alpha | \psi}|^2.
    \label{qpure}
\end{equation}

When expressed in physical phase-space variables, we use
\begin{equation}
    \alpha =
    \sqrt{\frac{m\omega}{2\hbar}}
    \left(x+\frac{i}{m\omega}p\right),
    \qquad
    \mathrm{d}^2\alpha
    =
    \frac{\mathrm{d}x\,\mathrm{d}p}{2\hbar}.
\end{equation}
Accordingly, the normalized phase-space density in the $(x,p)$ variables is
$Q(x,p)=Q(\alpha)/(2\hbar)$.
\subsubsection{Properties of the Husimi \texorpdfstring{$Q$}{Q} Function}
While the following properties are formulated for pure states, their generalization to mixed states via the density operator $\hat{\rho}$ is straightforward.

\begin{enumerate}[label=(\roman*)]
    \item \textbf{Non-negativity:} According to Eq.(\ref{qpure}), the $Q$ function is not only real
    but also non-negative everywhere in phase space, which is different from the Wigner function. 
    As a result, $Q(\alpha)$ may be interpreted as a probability density for the outcome of a simultaneous measurement of position and momentum.
  
    Despite being non-negative, it is still not a true probability distribution for two reasons. 
    Firstly, the coherent states are not orthogonal to each other as indicated by Eq.~\eqref{co_orth}. 
    Secondly, unlike the Wigner function, the marginal probability of $Q(\alpha)$ does not yield the exact quantum probability 
    $|\psi(x)|^2$ or $|\varphi(p)|^2$. To see this explicitly, 
    we integrate the $Q$ function over the momentum variable. 
    Using the convolution relationship with the Wigner function, 
    the spatial marginal distribution is obtained as:
    \begin{equation}
        \int Q(x,p) \mathrm{d}p = \sqrt{\frac{m\omega}{\pi\hbar}} \int |\psi(x')|^2 \exp\left[-\frac{m\omega}{\hbar}(x-x')^2\right] \mathrm{d}x'.
    \end{equation}
    This reveals that the marginal of the $Q$ function is the true spatial probability density $|\psi(x)|^2$ convoluted with a Gaussian 
    of width $\Delta x = \sqrt{\hbar / 2m\omega}$. Similarly, the momentum marginal is 
    also a Gaussian convolution of $|\varphi(p)|^2$ with width $\Delta p = \sqrt{\hbar m\omega / 2}$.
    This intrinsic ``blurring'' reflects the zero-point fluctuations of the coherent state probe.

    \item \textbf{Nonlinearity:} As a quadratic form of the state vector, the $Q$ function is nonlinear. 
    For a superposition $\ket{\psi} = c_1\ket{\psi_1} + c_2\ket{\psi_2}$, the distribution is:
    \begin{align}
        Q(\alpha) =& |c_1|^2 Q_1(\alpha) + |c_2|^2 Q_2(\alpha) \nonumber\\
        & + \frac{2}{\pi}\text{Re} \left( c_1 c_2^* \Braket{\alpha|\psi_1}\Braket{\psi_2|\alpha} \right).
    \end{align}
    The last term represents the quantum interference between the two components, albeit smoothed by the coherent state projection.

    \item \textbf{Normalization:} The $Q$ function is normalized over the complex $\alpha$-plane:
    \begin{equation}
        \int \mathrm{d}^2\alpha \, Q(\alpha) = 1,
    \end{equation}
    which follows directly from the resolution of unity in Eq.~\eqref{equ:Husimi_normalization}.

    \item \textbf{Boundedness and Analyticity:} The $Q$ function is bounded by $0 \leq Q(\alpha) \leq 1/\pi$. Because coherent states 
    are ``minimum uncertainty'' states, the $Q$ function is a smooth, real-analytic function that avoids the singularities often encountered in classical distributions or other quantum representations.

    \item \textbf{Invertibility and Numerical Sensitivity:} 
    While the global phase of $\ket{\psi}$ is lost, the mapping between the density operator $\hat{\rho}$ and $Q(\alpha)$ is mathematically bijective. 
    The matrix elements of $\hat{\rho}$ in the Fock basis can be retrieved via the following relation \cite{scully1997quantum}:
    \begin{equation}
        \Braket{n|\hat{\rho}|m} = \frac{\pi}{\sqrt{n!m!}} \left[ \frac{\partial^{n+m}}{\partial \alpha^{*n} \partial \alpha^m} \left( Q(\alpha, \alpha^*) e^{|\alpha|^2} \right) \right]_{\alpha=0}.
        \label{equ:Husimi_inverse}
    \end{equation}
    
    Although this result suggests that $Q(\alpha)$ contains the full information of the quantum state, reconstructing $\hat{\rho}$ from experimental 
    data is a notoriously \textit{ill-posed inverse problem}. The Weierstrass transform (the Gaussian convolution that generates $Q$) acts 
    as a powerful low-pass filter, exponentially suppressing the high-frequency interference terms that characterize purely quantum features. 
    In practice, computing the high-order derivatives required by Eq.~\eqref{equ:Husimi_inverse} is extremely sensitive to noise; even infinitesimal experimental uncertainties 
    in $Q(\alpha)$ are catastrophically amplified during the inversion process, making the fine-grained quantum information practically irretrievable.
    \item \textbf{Relationship with the Wigner Function:} In the dimensionless $\alpha$-plane, the $Q$ function can be obtained by convolving the Wigner function with a Gaussian filter:
    \begin{equation}
        Q(\alpha) = \frac{2}{\pi} \int \mathrm{d}^2\beta \, W(\beta) \exp(-2|\alpha-\beta|^2),
    \end{equation}
    where $W(\beta)$ denotes the Wigner function expressed in the same dimensionless phase-space coordinate. This ``smoothing'' explains 
    why $Q(\alpha)$ loses the fine-grained interference patterns (negative values) of the Wigner function.

    \item \textbf{Time Evolution:} The dynamical equation  for the $Q$ function can be derived from 
    the von Neumann equation as follows:
    \begin{equation}
        \frac{\partial Q(\alpha)}{\partial t}
=
-\frac{i}{\pi\hbar}
\bra{\alpha}[\hat{H},\hat{\rho}]\ket{\alpha}.
    \end{equation}
  In the semiclassical limit $\hbar \to 0$, the Husimi function converges to the classical ensemble distribution \cite{emamirad2013semiclassical} that satisfies the Liouville equation, 
  similar to the Moyal equation for the Wigner function (\ref{equ:Wigner_time}). 
  In this limit, the coherent state wave packets shrink to points in phase space (Dirac $\delta$-functions), and the quantum dynamics effectively reduce to classical Hamiltonian flow.
\end{enumerate}
\subsubsection{Example: The Quantum Harmonic Oscillator}
To further illustrate the $Q$ function, we calculate its distribution for the eigenstate $\ket{n}$ of a harmonic oscillator, 
which is also a Fock state. Using the expansion of the coherent state from Eq.~\eqref{equ:Husimi_superpostion}, the $Q$ function is given by:
\begin{equation}
    \begin{aligned}
        Q_n(\alpha) &= \frac{1}{\pi} \Braket{\alpha|n}\Braket{n|\alpha} \\
        &= \frac{e^{-|\alpha|^2} |\alpha|^{2n}}{\pi n!}.
    \end{aligned}
\end{equation}
Setting $m\omega = \hbar = 1$ for convenience, the radial distribution is solely a function of the phase-space distance $r = |\alpha|$.

\begin{figure}[htbp]
    \centering
    \includegraphics[width=0.9\columnwidth]{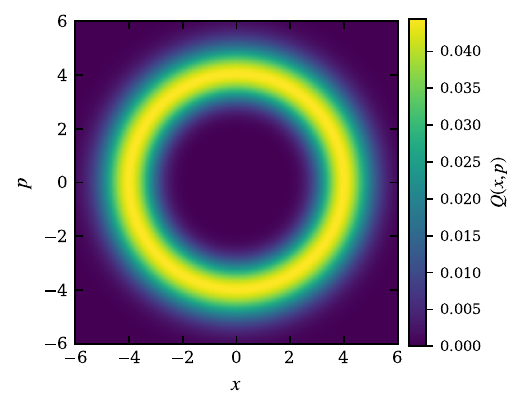}
    \caption{The Husimi distribution of the eighth energy eigenstate of a harmonic oscillator.}
    \label{fig:Husimi_harmonic}
\end{figure}

Fig.~\ref{fig:Husimi_harmonic} displays the Husimi $Q$ distribution for the $n=8$ state. A comparison between this result 
and the Wigner distribution in Fig.~\ref{fig:Wigner_harmonic} reveals several key insights:
\begin{enumerate}[label=(\roman*)]
\item The $Q$ function is strictly non-negative and lacks the rapid oscillations (interference fringes) characteristic of the Wigner function. 
This ``smoothness'' is a direct consequence of the Gaussian coarse-graining inherent in the Husimi representation. 
\item The distribution vanishes at the origin ($\alpha=0$). This stands in stark contrast to the Wigner function of the same state, which exhibits a strong (often negative) peak at the center. 

\item The distribution is concentrated along a circle in phase space. The radius of this circle, found by maximizing $Q_n(\alpha)$, is $|\alpha| = \sqrt{n}$. 
In the classical limit, this corresponds to the circular trajectory of an oscillator with energy $E = (n + 1/2)\hbar\omega$. Thus, while the Wigner function captures the intricate quantum interference of the state, 
the Husimi $Q$ function provides a more ``classical-like'' visualization, effectively mapping the quantum state onto the vicinity of its classical counterpart.
\end{enumerate}

\subsection{The Glauber--Sudarshan \texorpdfstring{$P$}{P} Function}
\subsubsection{Definition of the Glauber--Sudarshan \texorpdfstring{$P$}{P} Function}
The Glauber--Sudarshan $P$ function (hereafter referred to as the $P$ function) is another fundamental quasi-probability distribution 
in quantum phase space. It was  introduced independently by Sudarshan \cite{sudarshan1963equivalence}  
and by Glauber \cite{glauber1963coherent} in  1963. 
Unlike the Wigner and Husimi $Q$ functions, the $P$ function can exhibit severe singularities, 
such as derivatives of the Dirac $\delta$-function, and can take negative values for certain quantum states. 
Due to these features, it is an indispensable tool in quantum optics for identifying non-classical states of light.

The $P$ function is defined implicitly through the diagonal representation of the density operator in the coherent state basis:
\begin{equation}
    \hat{\rho} = \int \mathrm{d}^2\alpha \, P(\alpha) \ket{\alpha}\bra{\alpha},
    \label{equ:P_definition}
\end{equation}
where $\ket{\alpha}$ is the coherent state. This definition of the $P$ function differs fundamentally from those of the $W$ and $Q$ functions. 
Instead of providing an explicit functional form based on expectation values, Eq. (\ref{equ:P_definition}) expresses 
the density operator as a weighted expansion, where the $P$ function serves as the quasi-probability weight.

The existence of this representation is guaranteed by the overcompleteness of coherent states, although $P(\alpha)$ may not behave 
like a well-defined function in the classical sense, often requiring the framework of tempered distributions to handle its singularities.

\subsubsection{Properties of the Glauber--Sudarshan \texorpdfstring{$P$}{P} Function}
While the $P$ function is defined implicitly via the diagonal representation, it possesses several rigorous mathematical properties and explicit forms that highlight its unique role in phase space.

\begin{enumerate}
    \item \textbf{Explicit Representations and Normal Ordering:} 
    The $P$ function is closely associated with the expectation values of normally ordered operators. An operator $\hat{O}(\hat{a}^\dagger, \hat{a})$ 
    is in normal order if all creation operators $\hat{a}^\dagger$ precede all annihilation operators $\hat{a}$, i.e., $\hat{O} = \sum_{m,n} c_{mn} (\hat{a}^\dagger)^m \hat{a}^n$. 
    
    To construct an explicit form, we introduce the normally ordered Dirac $\delta$-function:
    \begin{align}
        &\delta(\alpha^* - \hat{a}^\dagger)\delta(\alpha - \hat{a}) \nonumber\\
        =& \frac{1}{\pi^2} \int \mathrm{d}^2\gamma \exp\left[-\gamma(\alpha^* - \hat{a}^\dagger)\right] \exp\left[\gamma^*(\alpha - \hat{a})\right].\nonumber\\
        \label{equ:delta}
    \end{align}
    This operator-valued distribution satisfies the following identity for any coherent state $\ket{\gamma}$:
    \begin{equation}
        \bra{\gamma}\delta(\alpha^* - \hat{a}^\dagger)\delta(\alpha - \hat{a})\ket{\gamma} = \delta^2(\alpha - \gamma).
    \end{equation}
    Furthermore, it acts as the basis for normally ordered moments:
    \begin{equation}
        (\hat{a}^\dagger)^m \hat{a}^n = \int \mathrm{d}^2\alpha \, \alpha^{*m} \alpha^n \delta(\alpha^* - \hat{a}^\dagger)\delta(\alpha - \hat{a}).
        \label{equ:P_norm}
    \end{equation}
    Using this, the $P$ function can be expressed as the trace of the density operator \cite{scully1997quantum}:
    \begin{equation}
        P(\alpha) =  \mathrm{Tr} \left[ \hat{\rho} \, \delta(\alpha^* - \hat{a}^\dagger)\delta(\alpha - \hat{a}) \right].
        \label{equ:P_tr}
    \end{equation}
    An alternative integral form, useful for practical calculations, is given by:
    \begin{equation}
        P(\alpha) = \frac{e^{|\alpha|^2}}{\pi^2} \int \mathrm{d}^2\gamma \bra{-\gamma} \hat{\rho} \ket{\gamma} e^{|\gamma|^2 - \gamma \alpha^* + \gamma^* \alpha}.
        \label{equ:P_definition2}
    \end{equation}

    \item \textbf{Normalization:} 
    The $P$ function is a normalized distribution. By integrating Eq. (\ref{equ:P_tr}) over the complex plane and utilizing the property in Eq. (\ref{equ:P_norm}), we obtain:
    \begin{align}
        &\int \mathrm{d}^2\alpha \, P(\alpha) \nonumber\\
        = & \mathrm{Tr}\left[ \hat{\rho} \int \mathrm{d}^2\alpha \, \delta(\alpha^* - \hat{a}^\dagger)\delta(\alpha - \hat{a}) \right] \nonumber\\
        = & \mathrm{Tr}[\hat{\rho}] = 1.
    \end{align}

    \item \textbf{Singularity and Non-classicality:} 
    A hallmark of the $P$ function is its potential for severe singularity and negative values, particularly for states lacking a classical counterpart.
    
    As a primary example, the $P$ function of a coherent state $\ket{\gamma}$ is:
    \begin{equation}
        P(\alpha) = \bra{\gamma} \delta(\alpha^* - \hat{a}^\dagger)\delta(\alpha - \hat{a}) \ket{\gamma} = \delta^2(\alpha - \gamma).
    \end{equation}
    Unlike the Gaussian distributions found in $W$ and $Q$ representations, the $P$ function of the ``most classical'' state is already a singularity (a $\delta$-peak).
    
    For a Fock state $\ket{n}$ or the $n$th energy eigenstate of a harmonic oscillator, the singularity becomes even more pronounced. Applying Eq. (\ref{equ:P_definition2}), we have
        \begin{align}
        P(\alpha)
        =& \frac{e^{|\alpha|^2}}{\pi^2}
        \int \mathrm{d}^2\gamma\,
        \braket{-\gamma|n}\braket{n|\gamma}
        e^{|\gamma|^2-\gamma\alpha^*+\gamma^*\alpha}
        \nonumber\\
        =& \frac{e^{|\alpha|^2}}{n!}
        \frac{\partial^{2n}}{\partial \alpha^n \partial \alpha^{*n}}
        \left(
        \frac{1}{\pi^2}
        \int \mathrm{d}^2\gamma\,
        e^{-\gamma\alpha^*+\gamma^*\alpha}
        \right)
        \nonumber\\
        =& \frac{e^{|\alpha|^2}}{n!}
        \frac{\partial^{2n}}{\partial \alpha^n \partial \alpha^{*n}}
        \delta^2(\alpha).
        \end{align}
The presence of higher-order derivatives of the $\delta$-function signifies that the $P$ function for number states is a highly non-classical 
distribution that cannot be interpreted as a standard function. This mirrors the difficulty of expanding pure Fock states as a statistical mixture of coherent states.

    \item \textbf{Transformation Hierarchy:} 
    The three quasi-probability functions are related via Gaussian convolutions, where $P$ acts as the most fundamental distribution:
     \begin{equation}
        W(\alpha) = \frac{2}{\pi} \int \mathrm{d}^2\gamma \, P(\gamma) e^{-2|\alpha-\gamma|^2},
    \end{equation}
    \begin{equation}
        Q(\alpha) = \frac{1}{\pi} \int \mathrm{d}^2\gamma \, P(\gamma) e^{-|\alpha-\gamma|^2}.
    \end{equation}
These relations show that the Wigner $W$ and Husimi $Q$ functions can be viewed as ``blurred'' versions of the $P$ function, which 
explains their generally better-behaved mathematical properties. Specifically, while the Wigner function may take negative values 
due to its derivation via a narrower Gaussian convolution, the Husimi $Q$ function is always non-negative, as it is obtained through convolution with a broader Gaussian kernel.  
\end{enumerate}
\subsubsection{Application: Single-Mode Thermal Equilibrium State (Photon Gas)}
A classic application of the $P$ function is the description of a single bosonic mode in thermal equilibrium. The density operator is given by:
\begin{equation}
    \hat{\rho} = \frac{\exp(-\beta\hat{\mathcal{H}})}{\mathrm{Tr}[\exp(-\beta\hat{\mathcal{H}})]},
\end{equation}
where $\beta = \frac{1}{k_\mathrm{B} T}$  and $\hat{\mathcal{H}} = \hbar\omega(\hat{a}^\dagger \hat{a} + \frac{1}{2})$. 
The matrix elements of $\hat{\rho}$ in the Fock basis are
\begin{equation}
    \bra{m}\hat{\rho}\ket{n}
    =
    \delta_{mn}
    \frac{\exp[-\beta\hbar\omega(n+1/2)]}{Z},
\end{equation}
where $Z=\mathrm{Tr}\!\left[e^{-\beta\hbar\omega(\hat{a}^\dagger\hat{a}+1/2)}\right]$ is the partition function.
It shows that the density matrix is diagonal in the Fock representation. The partition function 
can be computed and  has the following explicit form:
\begin{align}
   Z
=
\frac{e^{-\beta\hbar\omega/2}}{1-e^{-\beta\hbar\omega}}
=
e^{-\beta\hbar\omega/2}(\bar{n}+1).
\end{align}
Here $\bar{n}$ represents the average photon number, defined as $\bar{n} = \mathrm{Tr}[\hat{\rho}\hat{a}^\dagger\hat{a}] = (e^{\beta\hbar\omega}-1)^{-1}$.
Thus, the density operator can be written in terms of the average photon number:
\begin{equation}
    \hat{\rho} = \sum_{n=0}^{\infty} \ket{n}\bra{n} \frac{\bar{n}^{n}}{(\bar{n}+1)^{n+1}}.
\end{equation}

We can now calculate the $P$ function for this density operator using Eq. (\ref{equ:P_definition2}):
\begin{equation}
    \begin{aligned}
        P(\alpha) &= \frac{\mathrm{e}^{|\alpha|^2}}{\pi^2} \int \mathrm{d}^2\gamma \, 
        \bra{-\gamma} \hat{\rho} \ket{\gamma} \mathrm{e}^{|\gamma|^2 - \gamma \alpha^* + \gamma^* \alpha} \\
        &= \frac{1}{\pi \bar{n}} \exp\left( -\frac{|\alpha|^2}{\bar{n}} \right).
    \end{aligned}
\end{equation}
The result is evidently a Gaussian distribution in phase space. While the $P$ function is often not a regular function for pure nonclassical states, 
it is an effective method for describing statistical ensembles in quantum phase space.

It is meaningful to compare this result with the $Q$ and $W$ functions of the same thermal equilibrium state. 
The corresponding Wigner function in the dimensionless $\alpha$-plane is
\begin{equation}
    W(\alpha)
    =
    \frac{1}{\pi(\bar{n}+1/2)}
    \exp\left[
    -\frac{|\alpha|^2}{\bar{n}+1/2}
    \right],
\end{equation}
where we have utilized Eq. (\ref{equ:W_harmonic}) and the generating function of Laguerre polynomials. 
This is equivalent to increasing the variance in the $P$ function by $\frac{1}{2}$.

The corresponding $Q$ function is:
\begin{equation}
    \begin{aligned}
        Q(\alpha) &= \sum_{m=0}^{\infty} \frac{1}{\pi} |\Braket{\alpha|m}|^2 \frac{\bar{n}^{m}}{(\bar{n}+1)^{m+1}} \\
        &= \frac{1}{\pi (\bar{n} + 1)} \exp\left( -\frac{|\alpha|^2}{\bar{n} + 1} \right).
    \end{aligned}
\end{equation}
This is equivalent to increasing the average number of photons in the $P$ function by 1, resulting in a distribution that is smoother (broader) than the $P$ function.

This yields a remarkable conclusion: for a thermal equilibrium state in the canonical ensemble, the three quasi-probability distributions 
share the identical Gaussian form, differing only by a shift in the variance parameter. The $Q$ distribution is the smoothest (widest), 
followed by the $W$ distribution, while the $P$ distribution is the steepest (narrowest).

In the high-temperature limit (classical limit) where the average number of photons approaches infinity ($\bar{n} \to \infty$), 
the shifts of $1$ and $1/2$ become negligible. Consequently, these three distributions converge to the same classical Gibbs distribution, 
confirming the correspondence principle that quantum phase space distributions converge under classical conditions.

\subsection{From Characteristic Functions to Quasi-Probability Distributions}
All three  quasi-probability functions discussed above can be systematically generated from their respective characteristic functions. 
This unifies their definitions under the framework of Fourier transforms in the complex phase space.

\subsubsection{The P Function (Normal Ordering)}
The $P$ function is associated with the normally ordered characteristic function, defined as:
\begin{equation}
    C_\mathrm{N}(\beta) = \mathrm{Tr}\left[\hat{\rho} e^{\beta \hat{a}^\dagger} e^{-\beta^* \hat{a}}\right],
\end{equation}
where the creation operator $\hat{a}^\dagger$ stands to the left of the annihilation operator $\hat{a}$, satisfying the 
requirement of normal ordering. From this characteristic function, the $P$ function is recovered via a Fourier transform in the complex plane:
\begin{equation}
    \begin{aligned}
        P(\alpha) &= \mathrm{Tr}\left[\hat{\rho}\delta(\alpha^*-\hat{a}^\dagger)\delta(\alpha-\hat{a})\right] \\
        &= \mathrm{Tr}\left[\hat{\rho}\frac{1}{\pi^2}\int \mathrm{d}^2\beta \, e^{-\beta(\alpha^*-\hat{a}^\dagger)} e^{\beta^*(\alpha-\hat{a})}\right] \\
        &= \frac{1}{\pi^2} \int \mathrm{d}^2\beta \, e^{-\beta\alpha^*+\beta^*\alpha} C_\mathrm{N}(\beta),
    \end{aligned}
\end{equation}
where we have utilized the integral representation of the operator-valued $\delta$-function from Eq. (\ref{equ:delta}).

\subsubsection{The Q Function (Antinormal Ordering)}
The $Q$ function is generated from the antinormally ordered characteristic function:
\begin{equation}
    C_\mathrm{A}(\beta) = \mathrm{Tr}\left[\hat{\rho} e^{-\beta^* \hat{a}} e^{\beta \hat{a}^\dagger}\right],
\end{equation}
where the annihilation operator precedes the creation operator. The $Q$ function is obtained by the corresponding Fourier transform:
\begin{equation}
    Q(\alpha) = \frac{1}{\pi^2} \int \mathrm{d}^2\beta \, e^{-\beta\alpha^*+\beta^*\alpha} C_\mathrm{A}(\beta).
\end{equation}

\subsubsection{The Wigner Function (Weyl/Symmetric Ordering)}
The Wigner function corresponds to the Weyl (symmetric) ordered characteristic function:
\begin{equation}
    C_{\mathrm{W}}(\beta) = \mathrm{Tr}\left[\hat{\rho} \exp(-\beta^* \hat{a} + \beta \hat{a}^\dagger)\right],
\end{equation}
where $\hat{a}$ and $\hat{a}^\dagger$ appear symmetrically in the exponent (related to the displacement operator $\hat{D}(\beta)$). The Wigner function is generated by:
\begin{equation}
    W(\alpha) = \frac{1}{\pi^2} \int \mathrm{d}^2\beta \, e^{-\beta\alpha^*+\beta^*\alpha} C_{\mathrm{W}}(\beta).
\end{equation}

For a detailed derivation of these relations, one may refer to Ref. \cite{scully1997quantum}. This mathematical structure 
reveals a profound insight: the three distribution functions ($P, Q, W$), which appear quite different in their explicit forms 
and singularity properties, actually originate from the Fourier transforms of the characteristic functions corresponding 
to the three fundamental operator orderings (normal, antinormal, and symmetric).

\begin{table*}[!tbp]
\centering
\caption{Comparison of the three primary quasi-probability distributions in quantum phase space.}
\label{tab:quasi_prob_comparison}
\renewcommand{\arraystretch}{1.5}
\setlength{\tabcolsep}{10pt}
\begin{tabular}{lccc}
\hline\hline
\textbf{Property} & \textbf{P Function} & \textbf{Wigner Function (W)} & \textbf{Husimi Q Function} \\
\hline
\textbf{Operator Ordering} & Normal ($\hat{a}^\dagger \hat{a}$) & Symmetric (Weyl) & Antinormal ($\hat{a} \hat{a}^\dagger$) \\
\textbf{Definition} & Diagonal Rep. & Wigner Transform & $\frac{1}{\pi}\bra{\alpha}\hat{\rho}\ket{\alpha}$ \\
\textbf{Behavior} & Highly Singular & Oscillatory (can be $<0$) & Non-negative ($\ge 0$) \\
\textbf{Smoothing} & Steepest & Medium & Smoothest \\
\textbf{Ideal For} & Calculating normal-ordered moments & Quantum Interference & Semiclassical limit analysis \\
\hline\hline
\end{tabular}
\end{table*}

\subsubsection{Summary of Quasi-Probability Distributions}
To conclude this section, we summarize the key properties of the three distributions in Table \ref{tab:quasi_prob_comparison}.

\section{Construction of Basis in von Neumann's Planck Cells}
\label{sec:4}
\subsection{Historical Background and Introduction}

Well-known quantum phase space representations, such as the Wigner function, the Husimi $Q$ function, and 
the Glauber--Sudarshan $P$ function, all have certain drawbacks. As quasi-probability distributions, 
they do not satisfy all the requirements of a classical probability density. For instance, 
the Wigner function can take negative values, while the marginal distributions of the Husimi $Q$ function 
for a pure state do not recover the exact position or momentum densities $|\psi(x)|^2$ or $|\varphi(p)|^2$. 
Furthermore, the Glauber--Sudarshan $P$ function is highly singular and often unsuitable for describing pure states. 

Historically, von Neumann formulated a theoretical framework for quantum phase space before Wigner introduced the Wigner function.   
In his 1929 seminal paper \cite{neumann1929beweis,neumann2010}, as part of his work on the quantum ergodic 
theorem and the quantum $H$-theorem, von Neumann established a quantum phase space so that 
he could map a wave function to phase space. His basic idea was to divide classical phase space  
into Planck cells with a set of orthonormal and complete basis functions, each localized within a specific Planck cell. 

\begin{figure}[htbp]
    \centering
    \includegraphics[width=0.85\columnwidth]{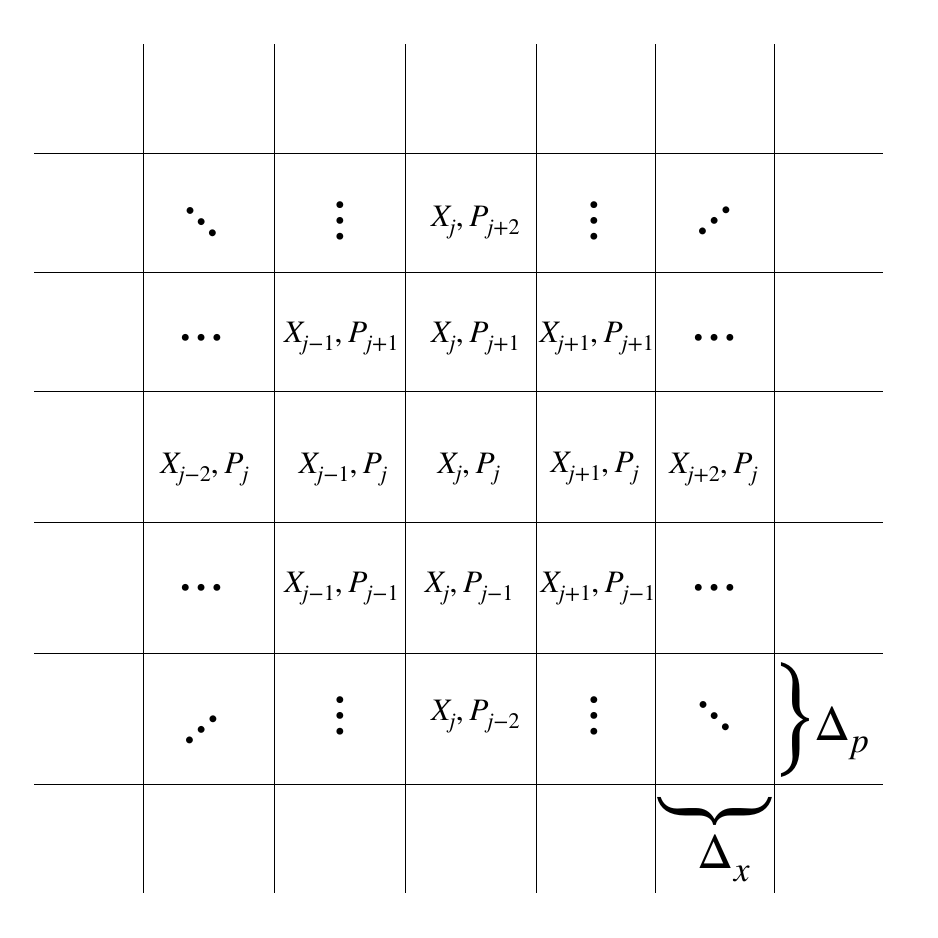}
    \caption{Quantum phase space with Planck cells. For each Planck cell, $\Delta_x\Delta_p=2\pi\hbar$.}
    \label{fig:cells}
\end{figure}

For simplicity, we first consider a one-dimensional system. In this case,  
von Neumann's quantum phase space is shown in Fig.~\ref{fig:cells}.  
Each Planck cell is centered at the point $(X_i, P_j)$ in phase space with  
coordinate $X_i = X_0 + i\Delta_x$ and momentum $P_j= P_0 + j\Delta_p$, where $i,j\in\mathbb{Z}$ label the position and momentum cells, respectively. 
$(X_0, P_0)$ denotes the origin of the phase-space lattice, and the phase-space area of each cell is fixed by $\Delta_x \Delta_p = 2\pi\hbar$. 
Each Planck cell is assigned a localized wave function $\ket{X_i,P_j}$. An ideal von Neumann basis $\{\ket{X_i,P_j}\}$
is required to satisfy the following orthonormality 
and completeness conditions:
\begin{align}
   & \Braket{X_i,P_j|X_{i'},P_{j'}} = \delta_{ii'}\delta_{jj'},\\
   & \sum_{i,j} \ket{X_i,P_j}\bra{X_i,P_j} = \hat{1}.
\end{align}
With this orthonormal basis, one can map a wave function $\ket{\psi}$ to phase space as 
\begin{equation}
\ket{\psi}=\sum_{i,j}a_{i,j}\ket{X_i,P_j}\,. 
\end{equation}
Unlike $P$, $Q$, and $W$ functions, this mapping to phase space is linear and unitary. 
As a result, $|a_{i,j}|^2$ is the probability of finding the system in the Planck cell labeled by $(i,j)$. \\

It is worthwhile to note that, with this basis set, von Neumann constructed a pair of operators as:
\begin{align}
\label{eq:mx}
    \hat{X} &= \sum_{i,j} X_i \ket{X_i,P_j}\bra{X_i,P_j},\\
    \hat{P} &= \sum_{i,j} P_j \ket{X_i,P_j}\bra{X_i,P_j}.
    \label{eq:mp}
\end{align}
They are analogous to the usual coordinate operator $\hat{x}$ 
and momentum operator $\hat{p}$, respectively, but differ crucially in that they are diagonal in the Planck-cell basis and commute with each other, $[\hat{X}, \hat{P}] = 0$. 
von Neumann called $\hat{X}$ the macroscopic coordinate operator and $\hat{P}$ the macroscopic momentum operator. 
Other macroscopic operators can be constructed via a similar procedure.
von Neumann utilized these concepts to bridge classical statistical physics and quantum statistical mechanics, 
succeeding in proving two foundational theorems: the quantum ergodic theorem and the quantum $H$-theorem \cite{neumann1929beweis}. 

However, it is not easy to compute and find these wave functions $\{\ket{X_i,P_j}\}$, which are localized
at Planck cells. In his 1929 paper \cite{neumann1929beweis,neumann2010}, 
von Neumann outlined a procedure to compute these wave functions by first assigning 
each Planck cell a Gaussian function and then orthonormalizing these Gaussian functions with 
the Gram--Schmidt method. For a large number of basis states, however, the Gram--Schmidt procedure becomes cumbersome and impractical. 

In the following sections, we will introduce two distinct methods for constructing the von Neumann basis in 
quantum phase space: the truncated basis \cite{wang2021microscope} and the quantum Wannier basis \cite{fang2018quantum, han2015entropy}.
Bourgain's nonperiodic construction \cite{bourgain1988remark}, which relaxes exact translational symmetry to 
improve localization, will be discussed separately in Section \ref{sec:5}.
\subsection{The Truncated Basis}

The truncated basis provides a straightforward and intuitive approach to constructing von Neumann's basis in quantum 
phase space \cite{wang2021microscope}. By truncating plane waves to the spatial interval associated with each Planck cell, we obtain the basis states:
\begin{equation}
    \ket{X_i,P_j} = \frac{1}{\sqrt{\Delta_x}} \int_{X_i - \frac{1}{2}\Delta_x}^{X_i + \frac{1}{2}\Delta_x} \mathrm{d}x\, e^{iP_j x/\hbar} \ket{x},
\end{equation}
where $\ket{x}$ is the eigenstate of the operator $\hat{x}$. 
This set of basis functions is strictly orthonormal and complete. By construction, we 
have $\braket{X_i,P_j|X_{i'},P_{j'}}=0$ when $i\neq i'$. When $i=i'$, we have 
\begin{equation}
    \begin{aligned}
        &\Braket{X_i,P_j|X_{i},P_{j'}} \\
               &= \frac{1}{\Delta_x}  \int_{X_i - \frac{\Delta_x}{2}}^{X_i + \frac{\Delta_x}{2}} \mathrm{d}x\, e^{\frac{i}{\hbar}(P_{j'} - P_j) x} =  \delta_{jj'},
    \end{aligned}
\end{equation}
where the integral vanishes for $j \neq j'$ because the phase factor 
$\exp[i(P_{j'}-P_j)x/\hbar]$ completes an integer number of periods over the interval of length $\Delta_x$. 
The completeness follows from the properties of the Fourier series: any function supported on the interval 
$[X_i - \frac{1}{2}\Delta_x, X_i + \frac{1}{2}\Delta_x]$ can be expanded unitarily in the plane wave basis $\{\frac{1}{\sqrt{\Delta_x}} e^{iP_j x/\hbar}\}$.

The functional forms of the truncated basis in position and momentum space are given by:
\begin{equation}
    \psi_{X_i,P_j}(x) = \frac{1}{\sqrt{\Delta_x}} \mathrm{rect}\left(\frac{x - X_i}{\Delta_x}\right) e^{iP_j x/\hbar},
\end{equation}
where $\mathrm{rect}(x)$ is the rectangular function:
\begin{equation}
    \mathrm{rect}(x) = \begin{cases} 
        1, & |x| \leq \frac{1}{2}, \\
        0, & |x| > \frac{1}{2}.
    \end{cases}
\end{equation}
In the momentum representation, the basis state is expressed as:
\begin{equation}
    \varphi_{X_i,P_j}(p) = \frac{1}{\sqrt{\Delta_p}} e^{-iX_i(p - P_j)/\hbar} \mathrm{sinc}\left(\frac{p - P_j}{\Delta_p}\right),
\end{equation}
where $\mathrm{sinc}(x) = \frac{\sin(\pi x)}{\pi x}$ is the normalized sinc function.

While the truncated basis is perfectly localized within a specific Planck cell in position space, it exhibits poor localization in momentum space. 
Specifically, the slow decay of the sinc function implies that the momentum variance is infinite, leading to $\sigma_x \sigma_p = \infty$, 
or equivalently $\sigma_x\sigma_k=\infty$ with $k=p/\hbar$. This behavior is a 
direct manifestation of the \textbf{Balian--Low theorem} \cite{balian1981principe}, which states that no complete 
and orthonormal basis set characterized by periodic translational invariance can be well-localized in both position and momentum space simultaneously.

\begin{figure}[htbp]
    \centering
    \includegraphics[width=0.85\columnwidth]{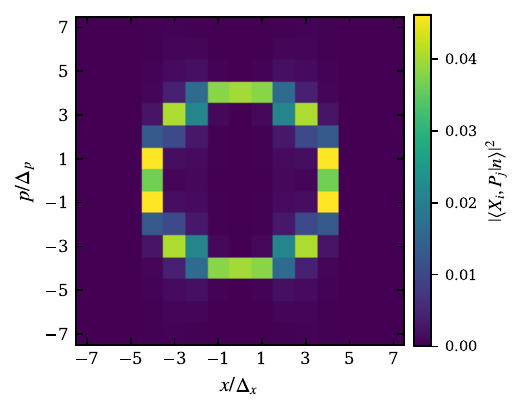} 
    \caption{Probability distribution of the eigenstate $\ket{n}$ ($n=50$) of a 
    harmonic oscillator in quantum phase space with the truncated basis. We set $X_0 = P_0 = 0$. 
    This distribution resembles the Husimi distribution in Fig.~\ref{fig:Husimi_harmonic}.}
    \label{fig:truncated_harmonic_50}
\end{figure}

As an illustration, we use the truncated basis to map the wave function for the $n$-th eigenstate $\ket{n}$ of a harmonic 
oscillator to phase space. The mapping is expressed as:
\begin{equation}
    \psi_n(x) =\braket{x|n}= \sum_{i,j} \Braket{X_i,P_j|n} \psi_{X_i,P_j}(x),
\end{equation}
where $\braket{X_i,P_j|n}$ represents the expansion coefficients (projections) of the state $\ket{n}$ onto the truncated 
basis $\ket{X_i,P_j}$. We configure the Planck cell dimensions as 
$\Delta_x = \sqrt{2\pi\hbar/m\omega}$ and $\Delta_p = \sqrt{2\pi\hbar m\omega}$. The resulting probability distribution $|\Braket{X_i,P_j|n}|^2$ 
over the Planck cells for $n=50$ is presented in Fig.~\ref{fig:truncated_harmonic_50}.  
In this figure, the probability distribution is seen primarily concentrated along the classical trajectory--specifically, 
a circle in the normalized phase space with radius proportional to $\sqrt{2n+1}$. 
This feature, also seen in the Husimi distribution in Fig.~\ref{fig:Husimi_harmonic}, is a direct 
demonstration of the quantum-classical correspondence. 

However, an evident asymmetry exists between the position and momentum directions; for instance, 
the statistical weights at the cells labeled by $(i,j)=(0,5)$ and $(5,0)$ are not identical. This asymmetry stems from 
the fundamental difference between the rectangular profile of the truncated basis 
in the position representation and its sinc function profile in the momentum representation. 
Unlike the Wigner and Husimi distributions, 
which often preserve the symmetry of the Hamiltonian's phase-space flow, the truncated 
basis imposes a specific spatial graining that breaks this rotational symmetry.

The truncated basis is computationally efficient and easily generalizable to higher dimensions. 
However, its lack of momentum localization restricts its utility in problems where phase-space compactness is critical. 
The Wannier basis introduced in the next section can achieve superior joint localization 
through numerical optimization.

\subsection{The Quantum Wannier Basis}
\label{sec:wannierb}

In the preceding sections, we briefly discussed von Neumann's attempt to construct an orthonormal basis for Planck cells. 
Before introducing the Wannier basis, we first review this construction in more detail. After partitioning the classical phase 
space into Planck cells, von Neumann assigned to these cells a set of Gaussian wave packets with width $\xi$:
\begin{equation}
\label{eq:dgaussian}
    g_{j_x,j_k}(x)
    =
    \left(\frac{1}{2\pi\xi^2}\right)^{1/4}
    \exp\left[
    -\frac{(x-j_xx_0)^2}{4\xi^2}
    +ij_kk_0x
    \right],
\end{equation}
where $j_x$ and $j_k$ are integers. The Gaussian wave packet $g_{j_x,j_k}(x)$ is localized around the Planck cell centered at $(j_xx_0,j_kk_0)$.

In the Wannier-basis construction, it is more convenient to work in the $(x,k)$ representation rather than the physical $(x,p)$ representation, where
\begin{equation}
    k=\frac{p}{\hbar}.
\end{equation}
Thus $k$ denotes the wave vector, and the physical momentum is recovered as $p=\hbar k$. In the $(x,k)$ representation, 
the phase-space cell area becomes dimensionless and is given by
\begin{equation}
    x_0k_0=2\pi.
\end{equation}
The set of Gaussian packets in Eq.~(\ref{eq:dgaussian}) forms a Gabor system \cite{gabor1946theory}, which is complete 
but not orthonormal. von Neumann originally suggested orthogonalizing these functions using the Gram--Schmidt method. 
However, this method has several significant drawbacks:
\begin{enumerate}
    \item \textit{Numerical instability}: The error tends to accumulate and increase during the orthogonalization process.
    \item \textit{Lack of translational symmetry}: The resulting basis functions do not preserve spatial translational invariance. 
    For a spatially uniform system, one expects the basis to behave identically across cells, which the Gram--Schmidt procedure destroys.
    \item \textit{Ordering sensitivity}: The outcome depends heavily on the order in which functions are orthogonalized, 
    and the resulting functions often bear little resemblance to the original Gaussian packets.
\end{enumerate}

It has been shown recently \cite{han2015entropy, fang2018quantum} that these drawbacks can be overcome by combining 
Kohn's orthogonalization \cite{kohn1973construction} and L\"owdin's method \cite{lowdin1950non}. 
The result is a complete orthonormal Wannier basis $\{\ket{w_{j_x,j_k}}\}$, which is translationally invariant along the $x$ direction. 
These Wannier functions $\ket{w_{j_x,j_k}}$ are centered and well localized at their respective cells $(j_xx_0,j_kk_0)$, and satisfy the orthonormality condition
\begin{equation}
    \Braket{w_{j_x',j_k'}|w_{j_x,j_k}}
    =
    \delta_{j_x'j_x}\delta_{j_k'j_k}.
\end{equation}
As a result, an arbitrary state $\ket{\psi}$ can be unitarily mapped into this discretized phase space:
\begin{equation}
    \ket{\psi}
    =
    \sum_{j_x,j_k}
    \Braket{w_{j_x,j_k}|\psi}\ket{w_{j_x,j_k}}.
\end{equation}
The coefficients $\Braket{w_{j_x,j_k}|\psi}$ give a \textit{true probability distribution} of quantum states in phase space, 
avoiding the negativity of the Wigner function and the over-smoothing of the Husimi $Q$ function. 
The computational methodology for constructing these functions is presented below.

We first review Kohn's procedure of orthogonalization \cite{kohn1973construction}, which may be unfamiliar to many readers. 
For simplicity, we consider a one-dimensional lattice composed of $N$ cells, each of length $a$. We impose periodic boundary 
conditions, as commonly adopted in solid-state physics. Let $g(x)$ be a function localized at $x=0$, and assume that $g(x)$ is not a Wannier function, namely
\begin{equation}
    \int dx\, g^*(x)g(x-ja)\neq 0
\end{equation}
for some nonzero integer $j$. A Gaussian function is such an example. Nevertheless, we can still construct Bloch functions from this non-Wannier function $g(x)$:
\begin{equation}
\label{eq:psim}
    \psi_m(x)
    =
    \frac{1}{\sqrt{N}}
    \sum_{j=0}^{N-1}
    e^{ijk_ma}g(x-ja),
\end{equation}
where $k_m$ is the wave vector in the first Brillouin zone. As a result of the periodic boundary condition, we have
\begin{equation}
    k_mNa=2m\pi,
    \qquad
    m\in\{0,1,\ldots,N-1\}.
\end{equation}
Even though $g(x)$ is not a Wannier function, it is straightforward to verify that the Bloch functions $\psi_m(x)$ are mutually orthogonal:
\begin{equation}
\label{eq:psinorm}
    \int_0^{Na}\mathrm{d}x\,\psi_m^*(x)\psi_{m'}(x)
    =
    c_m\delta_{m,m'}.
\end{equation}
In general, however, $c_m\neq 1$. We therefore normalize the Bloch wave functions by defining
\begin{equation}
    \chi_m(x)=\frac{\psi_m(x)}{\sqrt{c_m}},
\end{equation}
and use them to construct a localized wave function
\begin{equation}
    \mathfrak{g}(x)
    =
    \frac{1}{\sqrt{N}}
    \sum_{m=0}^{N-1}\chi_m(x).
\end{equation}
One can then show that $\mathfrak{g}(x)$ is a Wannier function:
\begin{align}
&\int_0^{Na} \mathrm{d}x\,\mathfrak{g}^*(x)\mathfrak{g}(x-ja)
\nonumber\\
=&
\frac{1}{N}
\int_0^{Na} \mathrm{d}x
\sum_{m,m'}
\chi_m^*(x)\chi_{m'}(x-ja)
\nonumber\\
=&
\frac{1}{N}
\int_0^{Na} \mathrm{d}x
\sum_{m,m'}
e^{-ijk_{m'}a}
\chi_m^*(x)\chi_{m'}(x)
\nonumber\\
=&\,\delta_{0,j}.
\end{align}
The above procedure is essentially an orthogonalization of the local wave functions $\{g(x-ja)\}$ into a set 
of Wannier functions $\{\mathfrak{g}(x-ja)\}$. This is Kohn's orthogonalization \cite{kohn1973construction}.

To numerically implement Kohn's orthogonalization, it is convenient to use Fourier transforms. For $g(x)$, we write
\begin{equation}
    g(x)
    =
    \frac{1}{\sqrt{Na}}
    \sum_n \tilde{g}(k_n)e^{ik_nx},
\end{equation}
where $k_n=2\pi n/(Na)$ due to the periodic boundary condition. Plugging this expansion into Eq.~(\ref{eq:psim}), we obtain
\begin{align}
\psi_m(x)
=&
\frac{1}{N\sqrt{a}}
\sum_{j=0}^{N-1}
e^{ijk_ma}
\sum_n \tilde{g}(k_n)e^{i(k_nx-jk_na)}
\nonumber\\
=&
\frac{1}{N\sqrt{a}}
\sum_n \tilde{g}(k_n)e^{ik_nx}
\sum_{j=0}^{N-1}e^{ij(k_m-k_n)a}
\nonumber\\
=&
\frac{1}{\sqrt{a}}
\sum_l \tilde{g}(k_{m+lN})e^{ik_{m+lN}x}.
\end{align}
This shows that the Bloch wave $\psi_m(x)$ is only related to a subset of Fourier coefficients $\{\tilde{g}(k_{m+lN})\}$ of the local function $g(x)$. According to Eq.~(\ref{eq:psinorm}), we find
\begin{equation}
\label{eq:gnorm}
    c_m
    =
    N\sum_l |\tilde{g}(k_{m+lN})|^2.
\end{equation}
Therefore, we can normalize $\psi_m(x)$ by choosing
\begin{equation}
    \tilde{\mathfrak{g}}(k_{m+lN})
    =
    \frac{\tilde{g}(k_{m+lN})}{\sqrt{c_m}}.
\end{equation}
With these $\tilde{\mathfrak{g}}(k_{m+lN})$ as Fourier coefficients, we obtain the Wannier function $\mathfrak{g}(x)$.

In summary, to numerically implement Kohn's orthogonalization, one only needs to Fourier transform $g(x)$, normalize its 
Fourier coefficients according to Eq.~(\ref{eq:gnorm}), and finally obtain the Wannier function $\mathfrak{g}(x)$ by inverse Fourier transform with the normalized coefficients.

However, Kohn's orthogonalization only orthonormalizes the Gaussian wave packets $\{g_{j_x,j_k}(x)\}$ along the $x$ direction. 
They remain nonorthogonal along the $k$ direction. The orthogonalization along the $k$ direction can be achieved using L\"owdin's 
method \cite{fang2018quantum}. The full orthogonalization procedure for $\{g_{j_x,j_k}(x)\}$ is as follows:
\begin{enumerate}
    \item \textit{Initialization}: Choose a set of local wave functions, such as the Gaussian packets $\{g_{j_x,j_k}(x)\}$. Set $j_x=0$ 
    and define $g_{j_k}(x)\equiv g_{0,j_k}(x)$. Compute its Fourier transform $\tilde{g}_{j_k}(k)$. Choosing $x_0=1$, $k_0=2\pi$, and $\xi=1/k_0$, 
    we obtain, up to a common normalization factor,
    \begin{equation}
        \tilde{g}_{j_k}(k)
        \propto
        \exp\left[
        -\left(\frac{k}{k_0}-j_k\right)^2
        \right].
    \end{equation}

    \item \textit{Vector construction}: To apply Kohn's orthogonalization, for a fixed $k\in[0,k_0)$, construct a column vector $f_{k,j_k}$ 
    whose $n$-th component is sampled from the momentum-space distribution:
    \begin{equation}
        f_{k,j_k}(n)
        =
        \tilde{g}_{j_k}(k+nk_0),
        \qquad n\in\mathbb{Z}.
    \end{equation}
    In numerical calculations, we introduce a cutoff $N_{\mathrm{cut}}$, restricting $-N_{\mathrm{cut}}\leq n\leq N_{\mathrm{cut}}$. 
    Vectors $f_{k,j_k}$ with different $j_k$ are not strictly orthogonal, although they approach orthogonality as $|j_k-j_k'|\to\infty$.

    \item \textit{L\"owdin orthogonalization}: We apply symmetric orthogonalization to the set $\{f_{k,j_k}\}$. We choose a cutoff $J_k$ such that $-J_k\leq j_k\leq J_k$.
    \begin{enumerate}
        \item Calculate the overlap matrix $S$:
        \begin{equation}
            S_{ij}
            =
            (f_{k,i},f_{k,j})
            =
            \sum_{n=-N_{\mathrm{cut}}}^{N_{\mathrm{cut}}}
            f_{k,i}^*(n)f_{k,j}(n),
        \end{equation}
        where $(\cdot,\cdot)$ denotes the vector inner product.

        \item Diagonalize the Hermitian overlap matrix:
        \begin{equation}
            S=U\Lambda U^\dagger,
        \end{equation}
        where $\Lambda$ is diagonal and $U$ is unitary.

        \item Compute the inverse square-root matrix:
        \begin{equation}
            S^{-1/2}=U\Lambda^{-1/2}U^\dagger.
        \end{equation}

        \item Transform the nonorthogonal vectors into an orthonormal set $\{u_{k,j}\}$:
        \begin{equation}
            u_{k,j}
            =
            \sum_{i=-J_k}^{J_k}
            f_{k,i}(S^{-1/2})_{ij}.
        \end{equation}
        We use the symmetric Fourier convention, and with this convention, 
        the orthogonalized function in $k$ space is defined as
        \begin{equation}
            \tilde{w}_{j_k}(k+nk_0)
            =
            \frac{u_{k,j_k}(n)}{\sqrt{k_0}}.
        \end{equation}
        For the choice $k_0=2\pi$, this reduces to $\tilde{w}_{j_k}(k+2n\pi)=u_{k,j_k}(n)/\sqrt{2\pi}$.
    \end{enumerate}

    \item \textit{Reconstruction}: Repeat steps 2 and 3 for every $k\in[0,k_0)$, discretized into $N_k$ points. Finally, perform the inverse 
    Fourier transform to obtain $w_{j_k}(x)$ in real space. The full Wannier basis is generated by spatial translation:
    \begin{equation}
        w_{j_x,j_k}(x)
        =
        w_{j_k}(x-j_xx_0).
    \end{equation}
    We denote $\ket{w_{j_x,j_k}}$ as $\ket{w_j}$ for brevity, where $j\equiv(j_x,j_k)$ is a composite cell index. The resulting basis satisfies
    \begin{equation}
        \Braket{w_{j_x',j_k'}|w_{j_x,j_k}}
        =
        \delta_{j_x'j_x}\delta_{j_k'j_k}.
    \end{equation}
    A detailed proof is provided in Appendix \ref{sec:Appendix_A}.
\end{enumerate}

For any given set of initial trial functions, L\"owdin orthogonalization yields a \textbf{unique} set of orthonormal vectors that are closest 
to the original set in the least-squares sense \cite{Lowdin1980}. This uniqueness and numerical stability resolve the drawbacks 
of the Gram--Schmidt method. Furthermore, as proved in Appendix \ref{sec:Appendix_A}, the resulting Wannier 
basis preserves spatial translational symmetry, providing a robust framework for phase-space quantization.

The Wannier basis constitutes a complete orthonormal basis in quantum phase space, provided that the cell dimensions satisfy $x_0k_0=2\pi$. 
The completeness originates from the fact that the Wannier basis is generated by the orthogonalization of translated Gaussian 
wave packets on Planck cells, namely from a complete Gabor system \cite{gabor1946theory}. Since the underlying Gabor system 
is complete under the critical-density condition \cite{balan2006density}, the Wannier basis derived from it inherits this completeness. 
The orthonormality condition
\begin{equation}
    \Braket{w_{j_x',j_k'}|w_{j_x,j_k}}
    =
    \delta_{j_x'j_x}\delta_{j_k'j_k}
\end{equation}
is rigorously proved in Appendix \ref{sec:Appendix_A}.

As a result, an arbitrary wave function $\ket{\psi}$ can be unitarily mapped into the Wannier basis:
\begin{equation}
    \ket{\psi}
    =\sum_j \ket{w_j}\Braket{w_j|\psi},
   \label{eq:wexp} 
\end{equation}
where $j\equiv(j_x,j_k)$ is the composite index of a Planck cell. The quantity
\begin{equation}
    \mathcal{P}_j
    =
    |\Braket{w_j|\psi}|^2
\end{equation}
represents the \textit{true probability} of finding the quantum state within the Planck cell labeled by $j$. Here the normalization 
is consistent with the symmetric Fourier convention introduced above. Unlike the Wigner $W$, Husimi $Q$, or Glauber--Sudarshan $P$ functions, 
this mapping is unitary and yields a strictly probabilistic distribution over the discretized phase space.

\begin{figure}[t]
    \centering
    \includegraphics[width=\columnwidth]{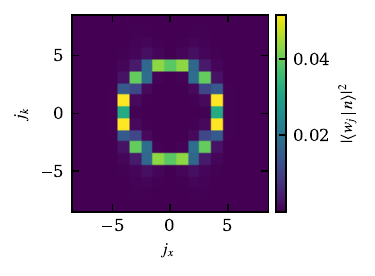}
    \caption{True probability distribution $|\langle w_j \mid n \rangle|^2$ of the $n=50$ harmonic-oscillator eigenstate 
    in the discretized Wannier phase space. The ring-like structure reflects the underlying classical orbit and is consistent with the phase-space distributions shown in Figs.~4 and 6.}
    \label{fig:wannier_phase_n50}
\end{figure}

We again use the harmonic oscillator to illustrate this construction. The results for the $n=50$ energy eigenstate 
are plotted in Fig.~\ref{fig:wannier_phase_n50}. In this figure, to align the circular symmetry of the harmonic oscillator's 
classical trajectory with our discretized phase space, where the Planck-cell dimensions $x_0$ and $k_0$ may differ, 
we rescale the plotting coordinates: the $x$ axis is compressed by a factor of $\sqrt{2\pi}$, while the $k$ axis is stretched by the same factor.

It is clear from Fig.~\ref{fig:wannier_phase_n50} that the distribution exhibits significant weight concentrated around 
the classical orbit, namely a circle in phase space with radius $R=\sqrt{(2n+1)/2\pi}$. We observe a slight asymmetry 
between the weights at the cells labeled by $(j_x,j_k)=(0,5)$ and $(5,0)$, which is a consequence of the basis construction 
lacking explicit rotational symmetry and exact translational symmetry in the $k$ direction. Compared with the truncated basis, 
the Wannier representation provides much sharper localization along the classical trajectory. This enhanced concentration 
demonstrates the superior phase-space resolution of the Wannier basis, which effectively avoids 
the over-spreading inherent in other discretized representations.

\begin{figure}[htbp]
    \centering
    \begin{subfigure}[t]{0.48\columnwidth}
        \centering
        \includegraphics[width=\columnwidth]{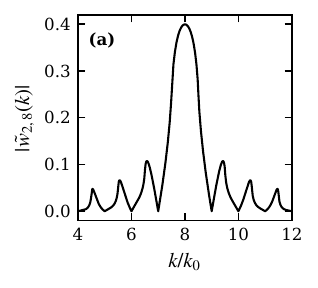}
        \label{fig:Wannier(a)}
    \end{subfigure}    
    \begin{subfigure}[t]{0.48\columnwidth}
        \centering
        \includegraphics[width=\columnwidth]{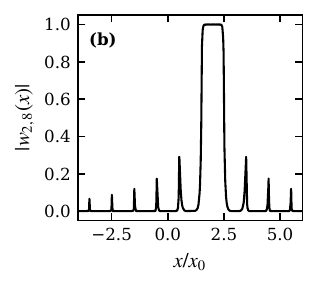}
        \label{fig:Wannier(b)}
    \end{subfigure}  
    \vfill
    \begin{subfigure}[t]{0.48\columnwidth}
        \centering
        \includegraphics[width=\columnwidth]{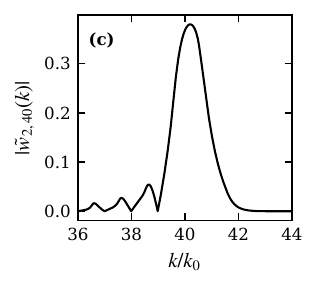}
        \label{fig:Wannier(c)}
    \end{subfigure}  
    \begin{subfigure}[t]{0.48\columnwidth}
        \centering
        \includegraphics[width=\columnwidth]{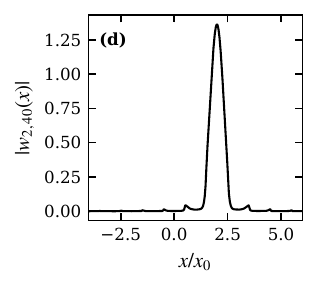}
        \label{fig:Wannier(d)}
    \end{subfigure}  
    \caption{Momentum- and position-space profiles of the Wannier basis functions $w_{2,8}$ and $w_{2,40}$. Panels (a) and (c) 
    show $|\tilde{w}_{2,8}(k)|$ and $|\tilde{w}_{2,40}(k)|$, while panels (b) and (d) show $|w_{2,8}(x)|$ and $|w_{2,40}(x)|$.}
    \label{fig:Wannier} 
\end{figure}

Some Wannier basis functions, such as $w_{2,8}$ and $w_{2,40}$, are demonstrated in Fig.~\ref{fig:Wannier} in both the $x$ and $k$ representations. 
These functions are real in the $k$ representation and complex in the $x$ representation. We observe that the Wannier functions are 
well localized around $(j_xx_0,j_kk_0)$ with oscillatory decaying tails. Notably, for small $j_k$, the basis shape deviates 
from a Gaussian and resembles a \textit{sinc} function. Conversely, for large $j_k$, the profile closely matches the initial Gaussian wave packet.

\begin{figure}[htbp]
    \centering
    \includegraphics[width=0.48\columnwidth]{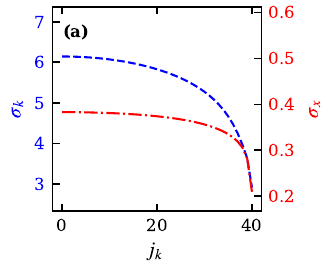}
    \hfill
    \includegraphics[width=0.48\columnwidth]{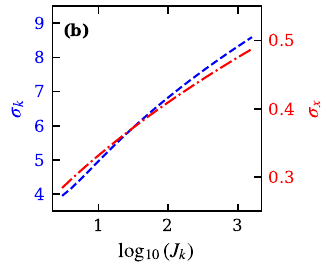}
    \caption{Localization properties of the Wannier basis. 
    (a) Dependence of the standard deviations $\sigma_x$ and $\sigma_k$ on the cell index $j_k$. 
    (b) Sub-logarithmic growth of $\sigma_x\sigma_k$ for the $j_k=0$ cell with respect to the cutoff $J_k$. 
    These two figures are recalculated with the same parameters as in \cite{fang2018quantum}.}
    \label{fig:walocal}
\end{figure}

We quantify the localization using the standard deviations $\sigma_x$ and $\sigma_k$. Due to spatial translational symmetry, 
it is sufficient to focus on the deviations of $w_{0,j_k}$. The results are plotted in Fig.~\ref{fig:walocal}, where both $\sigma_x$ 
and $\sigma_k$ reach their maxima at the center, $j_k=0$. As $j_k\to J_k$, the product $\sigma_x\sigma_k$ approaches 
the Gaussian limit of $1/2$. However, as the cutoff $J_k$ increases, the maximum value of $\sigma_x\sigma_k$, attained at $j_k=0$, 
grows at a sub-logarithmic rate. For a macroscopic system, such as $1\,\mathrm{m}^3$ of air where $J_k\approx 10^{34}$, 
the product $\sigma_x\sigma_k$ is only approximately $120$. This remains a remarkably small and manageable value 
compared with the truncated basis, for which the corresponding uncertainty product is already infinite in the ideal cutoff-free construction.

The ultimate divergence of $\sigma_x\sigma_k$ as $J_k\to\infty$ is a consequence of the \textit{Balian--Low theorem}, which 
prohibits a basis with strict translational symmetry from having finite variance in both representations simultaneously. 
A detailed proof is given in Appendix \ref{sec:Appendix_B}.

It is worth noting that the truncated basis can be viewed as a Wannier basis with poor localization in the $k$ direction.
\section{Applications of von Neumann's Quantum Phase Space}
\label{sec:App}
We have constructed two different sets of basis for Planck cells in  von Neumann's quantum phase space. 
In this section, we apply them in two different systems, respectively.

\begin{figure*}[t]
\centering
\includegraphics[width=0.8\textwidth]{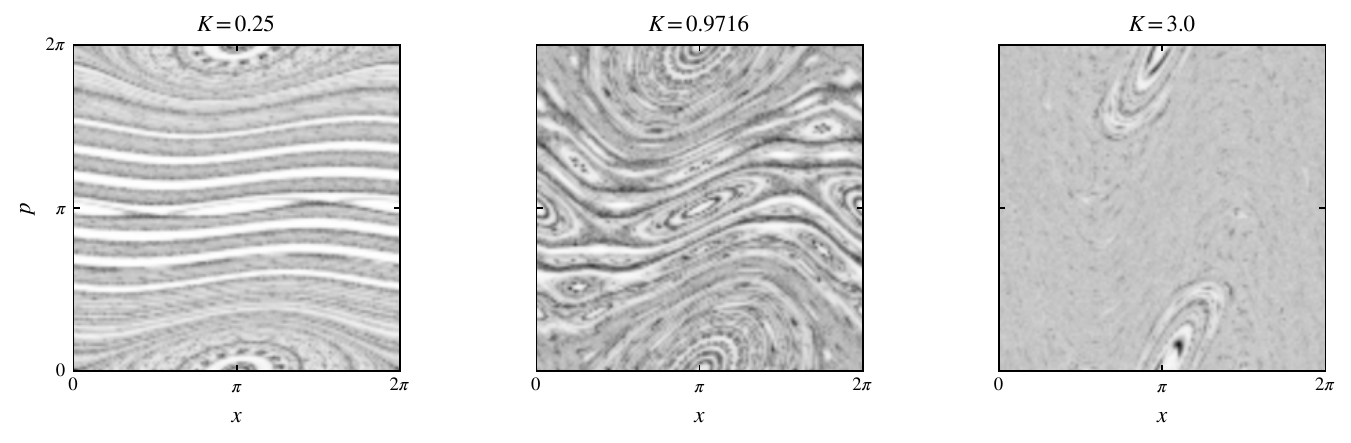} 
\caption{Quantum Poincar\'e sections of the kicked rotor for different kicking strengths. 
From left to right: $K=0.25$, $K=0.9716$, and $K=3.0$. 
The phase space is divided into an $L\times L$ grid of Planck cells with $L=128$ and 
$\hbar_{\mathrm{eff}}=2\pi/L^2$. 
The plots show the time-averaged phase-space probability distribution of an initial superposition state over 60 kicks.}
\label{fig:poincare_sections}
\end{figure*}

\subsection{Quantum Kicked Rotor}
\label{sec:QKR}

\subsubsection{The Model}

The quantum kicked rotor (QKR) is a paradigmatic model for investigating quantum chaos 
and the transition from integrable to ergodic dynamics. 
Its Hamiltonian can be obtained from the classical counterpart (\ref{eq:hckr}) by replacing $x$ 
with $\hat{x}$ and $p$ with $\hat{p}$:
\begin{equation}
\hat{H} = \frac{\hat{p}^2}{2} + K\cos \hat{x} \sum_{n=-\infty}^{\infty} \delta(t - n). 
\label{eq:dimensionless_H_quant}
\end{equation}
In this dimensionless representation, the commutation relation becomes 
$[\hat{x}, \hat{p}] = i\hbar_{\mathrm{eff}}$, where 
$\hbar_{\mathrm{eff}}=\hbar T/I$ is the effective Planck constant. 
In our numerical study, the Hilbert-space dimension is chosen as $N_{\mathrm H}=L^2$, which fixes
\begin{equation}
\hbar_{\mathrm{eff}} = \frac{2\pi}{N_{\mathrm H}}=\frac{2\pi}{L^2}.
\end{equation}

The quantum dynamics is described by the one-period evolution operator, known as the Floquet operator $\hat{U}$, 
which connects the state vectors immediately after consecutive kicks. Integrating the Schr\"odinger equation over one period yields:
\begin{equation}
\hat{U} = \hat{U}_{\mathrm{kick}} \hat{U}_{\mathrm{free}} 
= e^{-i \frac{K}{\hbar_{\mathrm{eff}}} \cos\hat{x}} 
e^{-i \frac{\hat{p}^2}{2\hbar_{\mathrm{eff}}}}.
\label{eq:Floquet_Op}
\end{equation}
The evolution consists of a free rotation term $e^{-i \hat{p}^2 / 2\hbar_{\mathrm{eff}}}$ followed by a kick term 
$e^{-iK\cos\hat{x}/\hbar_{\mathrm{eff}}}$. 

It is instructive to analyze the operator in the angular momentum basis $\ket{n}$, whose elements are eigenstates of the momentum operator $\hat{p}$ with eigenvalues $p_n=n\hbar_{\mathrm{eff}}$. In the $x$ representation, 
we have $\braket{x|n} = e^{inx}/\sqrt{2\pi}$. 
In terms of $\ket{n}$, the matrix elements of the Floquet operator can be expanded using Bessel functions of the first kind, $J_\nu(z)$:
\begin{equation}
\braket{m|\hat{U}|n} = (-i)^{m-n} J_{m-n}\left(\frac{K}{\hbar_{\mathrm{eff}}}\right) 
e^{-i \frac{\hbar_{\mathrm{eff}} n^2}{2}}.
\label{eq:Bessel_Expansion}
\end{equation}
This relation highlights the coupling between momentum states, where the transition amplitude between 
states $\ket{n}$ and $\ket{m}$ decays as the order of the Bessel function increases, governed by the parameter $K/\hbar_{\mathrm{eff}}$.
A more detailed derivation can be found in \cite{jiang2017quantum}.

\subsubsection{The Truncated Basis for QKR}

For the torus representation used here, we impose periodic boundary conditions in both the $x$ and $p$ directions and take the phase space to be 
$[0,2\pi)\times[0,2\pi)$. We divide this square into an $L\times L$ lattice, with $L$ 
being a large integer and $N_{\mathrm H}=L^2$. These small cells can be equivalently taken as Planck cells. 
To establish a quantum phase space for these Planck cells, we use the truncated Planck-cell basis, 
which was introduced in Ref.~\cite{jiang2017quantum} as the superposition-of-finite-momentum-eigenstates (SFME) basis.

We define the position and momentum translation operators acting on the momentum eigenstates $\ket{n}$ as:
\begin{align}
\hat{T}_x(X\Delta_x)\ket{n} &= \exp(-inX\Delta_x)\ket{n}, \label{eq:Tx} \\
\hat{T}_p(P\Delta_p)\ket{n} &= \ket{n + P\Delta_p / \hbar_{\mathrm{eff}}}, \label{eq:Tp}
\end{align}
where $X$ and $P$ are integers representing the lattice indices in phase space. The lattice constants are
\begin{equation}
    \Delta_x=\frac{2\pi}{L},
    \qquad
    \Delta_p=L\hbar_{\mathrm{eff}}=\frac{2\pi}{L},
\end{equation}
so that the unit cell area condition is satisfied:
\begin{equation}
    \Delta_x\Delta_p=2\pi\hbar_{\mathrm{eff}}.
\end{equation}

We construct a ``seed'' wave function by superposing $L$ consecutive momentum states:
\begin{equation}
\ket{0,0} = \frac{1}{\sqrt{L}} \sum_{\ell=0}^{L-1} \ket{\ell}.
\label{eq:seed}
\end{equation}
This state exhibits localization in both representations: it is a rectangular function in momentum space 
and, consequently, a periodic sinc-like function (Dirichlet kernel) in position space centered at $x=0$.
By applying the translation operators to this seed state, we generate the full orthonormal and complete basis set $\ket{X,P}$:
\begin{equation}
\ket{X,P} \equiv \hat{T}_x(X\Delta_x)\hat{T}_p(P\Delta_p)\ket{0,0}.
\label{eq:basis}
\end{equation}
For the periodic domain $x\in[0,2\pi)$, the position index runs over $X=0,1,\dots,L-1$. 
These basis states satisfy the orthonormality condition 
$\braket{X',P'|X,P}=\delta_{X'X}\delta_{P'P}$ and the completeness relation 
$\sum_{X,P}\ket{X,P}\bra{X,P}=\hat{1}$.

\subsubsection{Quantum Poincar\'e Sections of Kicked Rotor}

In Section~\ref{sec:2}, we have shown that the classical phase space can be used to illustrate 
the overall dynamical features of a system. In particular, for the classical kicked rotor, 
three Poincar\'e sections for different kick strengths $K$ are plotted in Fig.~\ref{fig:kicked_rotor_poincare}(a)--(c), from which 
one can clearly see how its dynamics changes from integrable to chaotic as the kick strength $K$ increases. 

With the truncated Planck-cell basis $\{\ket{X,P}\}$, we can also plot the Poincar\'e sections for 
the quantum dynamics of the kicked rotor. In our numerical computation, 
the initial state $\ket{\Psi(0)}$ is constructed as a superposition of sparsely 
distributed basis states $\ket{X_i,P_j}$, which is used to uniformly sample the phase space.
The system is evolved for a total duration of $60$ kicks. To replicate the stroboscopic nature of the classical Poincar\'e section, 
we accumulate the probability distribution at each time step. Crucially, to accommodate the toroidal topology of the phase space, 
we apply periodic boundary conditions in the momentum direction. This is achieved by summing the probabilities 
of Planck cells whose momentum indices $P$ differ by multiples of $L$ (corresponding to the $2\pi$ periodicity), as done in Ref.~\cite{jiang2017quantum}.

Figure~\ref{fig:poincare_sections} presents the resulting quantum Poincar\'e sections for three distinct dynamical regimes. 
For $K=0.25$ (left panel), the quantum distribution is mainly concentrated around the regular structures, showing qualitative 
agreement with the corresponding classical section in Fig.~\ref{fig:kicked_rotor_poincare}(a). 
As the kicking strength increases to the critical value $K=0.9716$ (middle panel), the distribution becomes more diffuse, 
reflecting the mixed phase-space structure also seen in Fig.~\ref{fig:kicked_rotor_poincare}(b). 
Finally, in the strongly chaotic regime with $K=3.0$ (right panel), the probability distribution spreads 
over a much larger portion of phase space, corresponding to the chaotic sea in Fig.~\ref{fig:kicked_rotor_poincare}(c).
A detailed discussion of this quantum-classical correspondence for the kicked rotor model can be found 
in Ref.~\cite{jiang2017quantum}. It is worthwhile to note that the dynamical transition seen in Fig.~\ref{fig:poincare_sections}
suggests that KAM-type structures also leave clear signatures in quantum phase-space representations.

The quantum Poincar\'e sections shown in Fig.~\ref{fig:poincare_sections} illustrate a key advantage of von Neumann's quantum phase space. 
Although Husimi-type representations can also produce smoothed stroboscopic phase-space portraits, 
as discussed in Appendix~\ref{app:husimi-qkr}, 
they are based on nonorthogonal coherent states and therefore do not yield probabilities over mutually exclusive Planck cells. 
Similarly, the Wigner function is sign-indefinite and typically contains rapid interference oscillations, 
making its direct visualization less suitable for this purpose. In contrast, the truncated Planck-cell basis gives a unitary projection onto an orthonormal Planck-cell basis, 
so that $|\braket{X,P|\psi}|^2$ is a genuine probability distribution on von Neumann's quantum phase space.
\subsection{Two-site Bose-Hubbard Model}
\label{sec:BoseHubbard}

So far, we have only considered quantum systems with infinite-dimensional Hilbert spaces. 
Here we construct a Planck-cell basis for a quantum system whose Hilbert space has finite dimension. 
For this purpose, we consider the two-site Bose-Hubbard model 
(for generalizations to three or more sites, refer to \cite{wang2021quantum}), which possesses 
a two-dimensional quantum phase space. The Hamiltonian of this model is given by
\begin{equation}
    \hat{H} 
    =
    -\dfrac{c_0}{2}
    \sum_{1\leq i,j\leq 2,\, i\neq j} 
    \hat{a}_i^\dagger \hat{a}_j
    +
    \dfrac{c}{2N_{\mathrm B}}
    \sum_{j=1}^{2} 
    \hat{a}_j^\dagger \hat{a}_j^\dagger \hat{a}_j \hat{a}_j ,
\end{equation}
where $\hat{a}_j^\dagger$ and $\hat{a}_j$ are the bosonic creation and annihilation operators for mode $j$, $c$ 
is the scaled interaction strength, and $N_{\mathrm B}$ is the total number of bosons. 
For this finite-dimensional system, a Wannier-type Planck-cell basis can be constructed in the Fock-number representation. 
The Fock states $\ket{N_1}$ satisfy
\begin{equation}
    \hat{N}_1\ket{N_1}
    =
    \hat{a}_1^\dagger\hat{a}_1\ket{N_1}
    =
    N_1\ket{N_1}.
\end{equation}

As $N_{\mathrm B}\to\infty$, the two-site Bose-Hubbard model admits a mean-field approximation. 
Using normalized mean-field amplitudes $a_1$ and $a_2$, with
\begin{equation}
    |a_1|^2+|a_2|^2=1,
\end{equation}
the corresponding mean-field Hamiltonian is
\begin{equation}
    H_{\mathrm{mf}}
    =
    -\dfrac{c_0}{2}
    \sum_{1\leq i,j\leq 2,\, i\neq j} 
    a_i^* a_j
    +
    \dfrac{c}{2}
    \sum_{j=1}^{2} |a_j|^4 .
\end{equation}
The behavior of the Bose-Hubbard model is primarily controlled by the particle number $N_{\mathrm B}$. 
When $N_{\mathrm B}$ is sufficiently large, the mean-field approximation serves as a classical counterpart and provides an accurate description of the model.

Since $a_1$ and $a_2$ are complex, we write
\begin{equation}
    a_1=|a_1|e^{i\theta_1},
    \qquad
    a_2=|a_2|e^{i\theta_2}.
\end{equation}
Due to the normalization condition $|a_1|^2+|a_2|^2=1$ and the irrelevance of the overall phase, only two independent variables remain. 
We choose the population fraction
\begin{equation}
    n_1=|a_1|^2
\end{equation}
and the relative phase
\begin{equation}
    \theta=\theta_1-\theta_2
\end{equation}
as the two mean-field phase-space coordinates. 
In the quantum description, $n_1$ is represented by $\hat{N}_1/N_{\mathrm B}$, while $\theta$ is represented by a discrete phase operator defined below.

The dimension of the Hilbert space is $N_{\mathrm B}+1$. 
For simplicity, we assume that
\begin{equation}
    N_{\mathrm B}+1=L^2,
\end{equation}
where $L$ is an integer. 
We then divide the number-phase phase space into $L\times L$ Planck cells. 
For each Planck cell, denoted by two indices $\ell,\vartheta=0,1,2,\cdots,L-1$, we construct the following Wannier-type Planck-cell basis:
\begin{equation}
    \ket{\ell, \vartheta} 
    =
    \dfrac{1}{\sqrt{L}}
    \sum_{m=0}^{L-1}
    \mathrm{e}^{i \frac{2\pi}{L}m\vartheta}
    \ket{m+\ell L}.
    \label{Bose_basis}
\end{equation}
Here $\ket{m+\ell L}$ is a Fock state of $\hat{N}_1$. 
It is straightforward to check that these basis states are orthonormal:
\begin{equation}
    \braket{\ell',\vartheta'|\ell,\vartheta}
    =
    \delta_{\ell'\ell}\delta_{\vartheta'\vartheta}.
\end{equation}
This construction is the one-dimensional analogue of the number-phase Planck-cell basis used for the three-site Bose-Hubbard model in Ref.~\cite{wang2021quantum}.

We now show that $\ket{\ell,\vartheta}$ is localized in the Planck cell labeled by $\ell$ and $\vartheta$. 
First, the state is uniformly supported on the number interval
\begin{equation}
    N_1=\ell L,\ell L+1,\ldots,\ell L+L-1.
\end{equation}
Therefore,
\begin{equation}
    \bra{\ell,\vartheta}\hat{N}_1\ket{\ell,\vartheta}
    =
    \ell L+\frac{L-1}{2}.
\end{equation}
Equivalently, the expectation value of the population fraction is
\begin{equation}
    \bra{\ell,\vartheta}\frac{\hat{N}_1}{N_{\mathrm B}}\ket{\ell,\vartheta}
    =
    \frac{\ell L+(L-1)/2}{N_{\mathrm B}}.
\end{equation}

Next, we define a discrete phase operator $\hat{\theta}$ through the Fourier-conjugate basis of $\hat{N}_1$:
\begin{equation}
    \hat{\theta}
    =
    \sum_{r=0}^{N_{\mathrm B}}
    \theta_r\ket{\theta_r}\bra{\theta_r},
\end{equation}
where $\ket{\theta_r}$ is the discrete Fourier transform of the Fock basis:
\begin{equation}
    \ket{\theta_r}
    =
    \dfrac{1}{\sqrt{N_{\mathrm B}+1}}
    \sum_{N_1=0}^{N_{\mathrm B}}
    \mathrm{e}^{iN_1\theta_r}\ket{N_1},
    \label{equ:Bose_Fourier}
\end{equation}
with
\begin{equation}
    \theta_r
    =
    \theta^{0}
    +
    \frac{2\pi r}{N_{\mathrm B}+1},
    \qquad
    r=0,1,\ldots,N_{\mathrm B}.
\end{equation}
Here $\theta^{0}$ is a phase origin satisfying $0\leq\theta^{0}\leq 2\pi/(N_{\mathrm B}+1)$. 
The orthonormality of $\ket{\theta_r}$ follows from the discrete Fourier transform:
\begin{equation}
    \Braket{\theta_{r'}|\theta_r}
    =
    \delta_{r'r}.
\end{equation}
The inverse Fourier transform is
\begin{equation}
    \ket{N_1}
    =
    \dfrac{1}{\sqrt{N_{\mathrm B}+1}}
    \sum_{r=0}^{N_{\mathrm B}}
    \mathrm{e}^{-iN_1\theta_r}\ket{\theta_r}.
\end{equation}
Using this relation, we can rewrite Eq.~\eqref{Bose_basis} as
\begin{equation}
    \ket{\ell,\vartheta}
    =
    \dfrac{1}{L^{3/2}}
    \sum_{r=0}^{N_{\mathrm B}}
    \dfrac{
    1-\mathrm{e}^{iL\left(\frac{2\pi\vartheta}{L}-\theta_r\right)}
    }{
    1-\mathrm{e}^{i\left(\frac{2\pi\vartheta}{L}-\theta_r\right)}
    }
    \mathrm{e}^{-iL\ell\theta_r}
    \ket{\theta_r}.
    \label{equ:Bose_basis2}
\end{equation}
This expression shows that the phase distribution is peaked around
\begin{equation}
    \theta=\frac{2\pi\vartheta}{L}.
\end{equation}

Because the Hilbert space is finite-dimensional, $\hat{N}_1$ and $\hat{\theta}$ do not satisfy an exact canonical commutation relation. 
Instead, their commutator contains boundary terms and depends on the choice of the phase origin $\theta^0$. 
This is the usual subtlety of number-phase conjugacy in a finite-dimensional Hilbert space. 
Nevertheless, the pair $(\hat{N}_1,\hat{\theta})$ provides an appropriate discrete number-phase representation for constructing Planck cells.

The expectation value of $\hat{\theta}$ under the basis state $\ket{\ell,\vartheta}$ can be calculated using Eq.~\eqref{equ:Bose_basis2}:
\begin{equation}
    \bra{\ell,\vartheta}\hat{\theta}\ket{\ell,\vartheta}
    =
    \dfrac{2\pi\vartheta}{L}
    +
    \dfrac{1}{L^3}
    \sum_{r=0}^{N_{\mathrm B}}
    \left|
    \dfrac{
    \sin\left(\frac{L\tilde{\theta}_r}{2}\right)
    }{
    \sin\left(\frac{\tilde{\theta}_r}{2}\right)
    }
    \right|^2
    \tilde{\theta}_r,
\end{equation}
where
\begin{equation}
    \tilde{\theta}_r
    =
    \theta_r-\frac{2\pi\vartheta}{L}.
\end{equation}
For large $L$, the second term in the phase expectation is approximately a constant independent of $\vartheta$, as shown in \cite{wang2021quantum}. 
Thus, the expectation values of $\hat{N}_1/N_{\mathrm B}$ and $\hat{\theta}$ scale linearly with $\ell$ and $\vartheta$, 
respectively, reflecting the physical locality of this basis set in the discrete number-phase phase space.

The localization of these Wannier-type states $\ket{\ell,\vartheta}$ can be quantified by their fluctuations in normalized particle number and phase. 
For the number direction, we obtain
\begin{equation}
    \Delta N_1
    =
    \sqrt{
    \langle \hat{N}_1^2\rangle
    -
    \langle \hat{N}_1\rangle^2
    }
    =
    \sqrt{\frac{L^2-1}{12}},
\end{equation}
and therefore the normalized particle-number fluctuation is
\begin{equation}
    \delta n_1
    =
    \frac{\Delta N_1}{N_{\mathrm B}}
    =
    \frac{1}{\sqrt{12N_{\mathrm B}}}
    \sim
    \frac{1}{\sqrt{12}L}.
\end{equation}
For the phase direction, we define
\begin{equation}
    \delta\theta
    =
    \sqrt{
    \langle \hat{\theta}^2\rangle
    -
    \langle \hat{\theta}\rangle^2
    }.
\end{equation}
Using Eq.~\eqref{equ:Bose_basis2}, this can be written as
\begin{equation}
    \begin{aligned}
        \delta\theta
        &=
        \sqrt{
        \frac{1}{L^3}
        \sum_{r=0}^{N_{\mathrm B}}
        \left|
        \frac{
        \sin\left(\frac{L\tilde{\theta}_r}{2}\right)
        }{
        \sin\left(\frac{\tilde{\theta}_r}{2}\right)
        }
        \right|^2
        \tilde{\theta}_r^2
        -
        C_\theta^2
        } \\
        &\sim \frac{1}{\sqrt{L}},
    \end{aligned}
\end{equation}
where $C_\theta$ is the constant expectation shift. 
The above results show that the basis is well localized in both population fraction and relative phase as $L\to\infty$. 
This construction provides a clear realization of Planck cells in number-phase space for finite-dimensional Bose-Hubbard systems.

\subsection{Entropy for Quantum States}
The Wannier basis--and other orthonormal bases constructed from Planck cells--offers a key advantage over the 
quasi-probability distributions such as the Husimi $Q$ function: it yields a genuine probability distribution
over phase space. This advantage was fully explored by von Neumann in 1929 to define 
an entropy for quantum states as part of his proof of the  
quantum H theorem~\cite{neumann1929beweis,neumann2010}. 

By 1927, von Neumann had already introduced an entropy for quantum states~\cite{vN1927}, 
now widely known as the von Neumann entropy. However, as he clearly noted in his 1929 
paper~\cite{neumann2010},  this entropy is unsuitable for the quantum H-theorem 
because it was ``computed from the perspective of an observer who can carry out all measurements
that are possible in principle'' and, consequently, vanishes for all pure states. 
Von Neumann recognized that for the scenarios to which  the quantum H theorem applies, 
``the observer can measure only macroscopically''.  To define a quantum entropy appropriate for these situations, von Neumann 
used a coarse-grained probability distribution 
derived from his theory of quantum phase space. 
In his view, coarse graining was necessary because a single macroscopic state comprises 
a collection of many microscopic quantum states, each associated with Planck cells.

Interestingly, von Neumann acknowledged in his 1929 paper that this definition of a ``new'' 
quantum entropy was borrowed from unpublished results of Wigner. 
To distinguish the two quantum entropies, we refer to the 1927 entropy as the von Neumann (vN) entropy, 
which is always zero for pure states of the full system, and the 1929 entropy as the Wigner-von Neumann (WvN) entropy, 
which is nonzero for almost all pure states.

The fates of these two quantum entropies are drastically different. The vN entropy 
is well known and finds wide applications. However, before its use in quantifying quantum entanglement, 
the vN entropy was often incorrectly identified with thermodynamic entropy--a mistake we will clarify later.
In contrast, the WvN entropy remains largely obscure and had never been developed or applied before one of the authors 
revisited it in 2015. One possible technical barrier is that the number of microscopic quantum states corresponding 
to a single macroscopic state cannot be clearly defined,  making the WvN entropy uncomputable in practice.

In 2015, the WvN entropy was re-examined in Ref.\cite{han2015entropy}, where coarse graining was found to be 
unnecessary. Once the Wannier basis $\ket{w_j}$ is established, one can project  a wave function $\ket{\psi}$
onto phase space as in Eq.(\ref{eq:wexp}) and define the WvN entropy as 
\begin{equation}
    S_w(\ket{\psi})
    =
    -\sum_j p_j\ln p_j,
    \qquad
    p_j=|\braket{w_j|\psi}|^2 .
    \label{eq:WvN_entropy}
\end{equation}
Since the Wannier basis can be efficiently computed numerically, the WvN entropy can also 
be efficiently calculated for a given quantum state $\ket{\psi}$. Furthermore,
with this definition, one can prove an inequality similar to that established by von Neumann 
for the  quantum H theorem. 

However, beyond spatial degrees of freedom, quantum systems also possess spin. It remains unclear 
how to define the WvN entropy in such cases, as the notion of a Wannier basis for spin degrees 
of freedom is not well defined. Ultimately, one realizes that for a given quantum system, 
among the infinitely many sets of orthonormal bases, only a few carry physical meaning 
or correspond to realistic physical measurements. In von Neumann's view, these 
are scenarios where  ``the observer can measure only macroscopically''.

 The Wannier basis is clearly physical for at least two reasons. First, when particles interact with each other, 
 although the interaction typically depends only on positions, momenta also play a significant role in determining the interaction outcome. 
 For example, in materials, when the kinetic energy of electrons dominates, we approximately have a system of free electrons; 
 when the interaction dominates, we have a strongly correlated material.
Second, the Wannier basis can be viewed as corresponding to measurements of macroscopic position and momentum 
defined in Eqs.~(\ref{eq:mx},\ref{eq:mp}). In fact, this kind of measurement has been realized using the cloud chamber, 
where the deterministic trajectories indicate that particles in the chamber possess approximately definite positions and momenta simultaneously at any given moment.

For spin systems, one can similarly select a set of complete orthonormal bases that are physical in the sense 
discussed above, obtain a probability distribution over this basis, and define the WvN entropy for a quantum state.

 This generalized WvN entropy was discussed and compared to the well-known vN entropy 
 in Ref.~\cite{Hu2019PRE}. There are two key differences between them. The first, mentioned earlier, is that the WvN entropy 
 is generally non-zero for pure quantum states, whereas the vN entropy always vanishes for pure states.
  
The second difference relates to extensiveness.  It was found that the WvN entropy is extensive--that is, 
 the WvN entropy of an entire system equals the sum of the entropies of its subsystems. 
 This stands in stark contrast to the vN entropy, which is non-extensive. For example, 
 when an isolated system is in an entangled pure state, the vN entropy of the whole system 
 is zero while the vN entropies of its subsystems are nonzero.

However, there is  one particular case in which the WvN entropy and the vN entropy are almost identical. 
Consider an isolated  system in an arbitrary entangled pure quantum state, divided  into two subsystems, $S$ and $B$,  
with $B$ much larger than $S$.  For the subsystem $S$, the vN and WvN entropies  
are almost the same~\cite{Hu2019PRE}. This coincidence carries an important implication, as discussed below.

With these developments, our understanding of these two quantum entropies needs to be updated. 
Even though the mathematical form of the vN entropy has remained unchanged, its meaning  
is no longer what von Neumann had originally in mind. In modern usage, the vN entropy is 
often interpreted as an entanglement entropy when applied to reduced density matrices. 
For an isolated system in a pure state, however, it should not be directly identified with thermodynamic entropy.
On the other hand, the WvN entropy has a slightly different mathematical form as it does not use 
coarse graining and has been generalized to spins, 
but it is still the entropy related to thermodynamics just as von Neumann had originally intended. 
In particular, when a system is in thermal equilibrium, the WvN entropy equals the thermodynamic entropy. 

A common shorthand in the literature is to identify the vN entropy with thermodynamic entropy. 
This identification can be misleading, especially for isolated pure states. Nevertheless, 
it does not lead to any unphysical consequences, owing to the coincidence mentioned earlier: 
for a small subsystem $S$, the vN entropy and the WvN entropy are nearly identical in values, even though they originate from different conceptual frameworks.


\subsection{Relation to Time--Frequency Analysis}

The construction of the quantum Wannier basis in quantum phase space is closely related to a broader class of problems 
in time--frequency analysis. In signal processing, one often seeks a representation that localizes a signal simultaneously 
in time and frequency. This is mathematically analogous to the problem of constructing quantum states localized 
in both position and momentum. From this perspective, the Planck-cell basis in von Neumann's quantum 
phase space can be viewed as a quantum-mechanical counterpart of localized time--frequency representations.

A standard example is the short-time Fourier transform (STFT), or more generally Gabor analysis. In this approach, 
a fixed window function is translated in time and modulated in frequency, producing a representation on a regular time--frequency lattice. 
Its main advantage is that the resolution is uniform over the whole time--frequency plane. However, 
when one demands both orthonormality and good localization in the two conjugate variables, a fundamental obstruction 
appears: the Balian--Low theorem states that an orthonormal Gabor basis with exact lattice translational symmetry 
cannot have finite second moments in both time and frequency \cite{gabor1946theory,balian1981principe,low1985complete}. 
This obstruction is directly analogous to the difficulty of constructing an orthonormal basis localized in each Planck cell of quantum phase space.

Wavelet analysis provides a different strategy. Instead of using translations and frequency modulations of a fixed window, 
wavelet bases use translations and dilations of a mother wavelet \cite{daubechies1992ten}. This gives a multiresolution representation: 
low-frequency components are described with better frequency resolution and poorer time resolution, while high-frequency components 
are described with better time resolution and poorer frequency resolution. Such a structure is especially useful for signals 
with edges, bursts, singularities, or structures occurring on multiple scales. However, the price is that the resolution is not uniform across the time--frequency plane.

The quantum Wannier or von Neumann phase-space basis is closer in spirit to Gabor analysis than to wavelet analysis. 
The two Fourier-conjugate variables, such as position and momentum, are discretized into Planck cells, and one seeks a localized 
basis state associated with each cell. The commonality among the Wannier construction, STFT, Gabor analysis, 
and wavelet analysis lies in the shared goal of representing a state or signal in a localized phase-space or time--frequency form. 
The essential difference is that the Wannier construction emphasizes an orthonormal and complete Planck-cell basis. 
As a result, the mapping from the Hilbert space to the discretized phase space is unitary, and the coefficients 
$|\langle w_j|\psi\rangle|^2$ define a genuine discrete probability distribution without loss of information. 
This feature distinguishes the Wannier phase-space representation from overcomplete, smoothed, 
or phase-insensitive time--frequency representations such as the STFT spectrogram or the Husimi function.

This connection also explains why the Balian--Low theorem and Bourgain's nonperiodic construction are relevant to quantum phase space. 
The former reveals the obstruction caused by imposing exact lattice translational symmetry together with orthonormality and strong localization, 
while the latter shows that relaxing strict periodicity can improve simultaneous localization in conjugate variables. 
Thus, the construction of quantum phase-space bases is not only a problem in the foundations of quantum mechanics, 
but also part of the broader mathematical theme of localized representations in harmonic analysis and signal processing. 
This observation motivates the Bourgain construction discussed in Section \ref{sec:5}.
\section{Bourgain's Nonperiodic Basis}
\label{sec:5}

The truncated basis is clearly not well localized. The quantum Wannier basis constructed in Section \ref{sec:wannierb} 
is much better localized than the truncated basis. However, because it preserves exact spatial translational symmetry, 
the Balian--Low theorem implies that in the infinite translationally invariant limit it cannot have finite second moments in both $x$ and $k$ \cite{balian1981principe}. 
The Balian--Low theorem states that any Gabor basis $\{g_{j_x,j_k}\}$ 
with phase-space translational invariance,
\begin{equation}
    g_{j_x,j_k}(x)=\mathrm{e}^{ij_k k_0 x}g(x-j_x x_0),
\end{equation}
with $x_0 k_0=2\pi$, must satisfy  
\begin{equation}
    \sigma_x(g)\sigma_k(g)=\infty .
\end{equation}
Later, Battle gave a simpler proof of the Balian--Low theorem in \cite{battle1988heisenberg}.

We may wonder how well localized a basis in phase space can be after removing translational symmetry. 
Steger proved a no-go theorem in his unpublished work that $L^2(\mathbb{R})$ does not 
permit a set of orthonormal basis functions $\{g_{j_x,j_k}\}$ satisfying
\begin{multline}
    \sup_{j_x,j_k} \bigg[
    \int (x-\bar{x}_{g_{j_x,j_k}})^{2+\varepsilon}|g_{j_x,j_k}(x)|^2 \mathrm{d}x\\
    +\int (k-\bar{k}_{\tilde{g}_{j_x,j_k}})^{2+\varepsilon}
    |\tilde{g}_{j_x,j_k}(k)|^2 \mathrm{d}k
    \bigg] < \infty,
\end{multline}
where $\varepsilon>0$ and $\tilde{g}_{j_x,j_k}(k)$ denotes the Fourier transform of $g_{j_x,j_k}(x)$. 
This theorem prohibits any basis from being uniformly localized in the sense of the $(2+\varepsilon)$-th central moment.

In Ref. \cite{bourgain1988remark}, Bourgain suggested a method 
for constructing an orthonormal basis $\{g_j\}$ in quantum phase space, reaching 
the lower bound restricted by Steger's no-go theorem. This basis satisfies
\begin{equation}
    \sup_{j} \bigg[
    \int (x-\bar{x}_{g_{j}})^{2}|g_{j}(x)|^2 \mathrm{d}x
    +
    \int (k-\bar{k}_{\tilde{g}_{j}})^{2}|\tilde{g}_{j}(k)|^2 \mathrm{d}k
    \bigg] < \infty .
\end{equation}
Such a basis is nonperiodic in phase space, and thus evades the restriction of the Balian--Low theorem.
Moreover, each element of this basis can be made arbitrarily close to a Gaussian wave packet with minimum uncertainty:
\begin{equation}
    \sigma_x(g_j)^2
    :=
    \int (x-\bar{x}_{g_j})^{2}|g_j(x)|^2 \mathrm{d}x
    < \dfrac{1}{2}+\varepsilon,
\end{equation}
and
\begin{equation}
    \sigma_k(g_j)^2
    :=
    \int (k-\bar{k}_{\tilde{g}_j})^{2}|\tilde{g}_j(k)|^2 \mathrm{d}k
    < \dfrac{1}{2}+\varepsilon,
\end{equation}
where $\varepsilon>0$.
Bourgain's construction is elegant, yet somewhat hard to comprehend. 
We will describe his construction below in a way that is hopefully easier to understand.

Bourgain's construction consists of two building blocks. One is a set of orthonormal wave functions  
$\{r_{\alpha,\beta}\}$, which are well localized on sites of a square lattice $T\times T$ ($T$ is a sufficiently large integer). 
The other is a dense sequence of compactly supported smooth functions $\{f_j\}$. One of 
the functions in the dense sequence is selected to linearly mix with $\{r_{\alpha,\beta}\}$ to 
construct another set of orthonormal functions $\{b_{\alpha,\beta}\}$. The procedure is repeated
infinitely many times so that the whole phase space is fully covered by the lattices, and 
the orthonormal functions $\{b_{\alpha,\beta}\}$ built in each step are combined to form
a complete basis for quantum phase space. When we mix the functions in the dense sequence 
with the well localized lattice functions, the weights are heavily tilted towards the latter. As a result, 
the final basis functions are all well localized. 
 
In the following we first introduce these two building blocks and then describe the whole procedure 
for constructing the basis functions. For simplicity, and also because the results are purely mathematical, 
the variables presented below are all dimensionless. 

\subsection{Mollifier and Dense Sequence of Functions}

\begin{figure}[htbp]
    \centering
    \begin{subfigure}[t]{0.45\columnwidth}
        \centering
        \includegraphics[width=0.9\columnwidth]{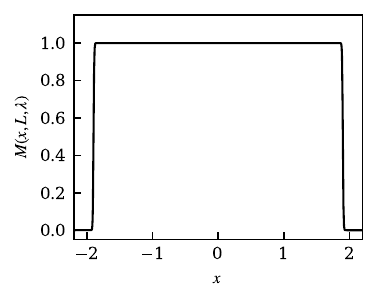}
        \captionsetup{justification=centering}
        \subcaption{$M(x,L,\lambda)$}
        \label{fig:Bourgain_mollifier1}
    \end{subfigure}
    \hfill
    \begin{subfigure}[t]{0.45\columnwidth}
        \centering
        \includegraphics[width=0.9\columnwidth]{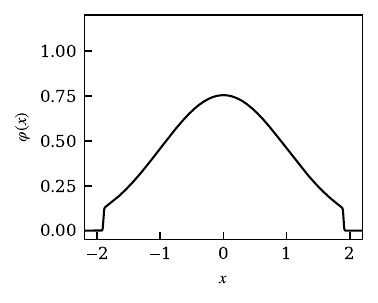}
        \captionsetup{justification=centering}
        \subcaption{$\varphi(x)$}
        \label{fig:Bourgain_mollifier2}
    \end{subfigure}
    \captionsetup{justification=centering}
    \caption{(a) The mollifier function $M(x,L,\lambda)$; (b) the truncated Gaussian function $\varphi(x)$. Here $L=4$ and $\lambda=0.9$.}
    \label{fig:Bourgain_mollifier}
\end{figure}

There are many ways to construct a mollifier function. One simple way is to utilize the $C^\infty$ function $F(x)$:
\begin{equation}
    F(x)=
    \begin{cases}
    \mathrm{e}^{-\frac{1}{x}}, & x>0, \\
    0, & x\leq 0.
    \end{cases}
\end{equation}
The mollifier $M(x,L,\lambda)$ can then be constructed from $F(x)$ as
\begin{equation}
    M(x,L,\lambda)
    =
    \dfrac{F(\frac{L}{2}-|x|)}
    {F(\frac{L}{2}-|x|)+F(|x|-\lambda\frac{L}{2})}.
\end{equation}
As shown in Fig.~\ref{fig:Bourgain_mollifier}(a), the parameter $L$ is the width of the function and $\lambda$ 
dictates how fast the function changes from zero to one. More precisely, $M(x,L,\lambda)=1$ 
when $|x|\leq \lambda L/2$ and equals $0$ when $|x|\geq L/2$. 
Thus, $M(x,L,\lambda)$ is a smooth function compactly supported in $[-L/2,L/2]$. 

The Gaussian functions in Eq.~(\ref{eq:dgaussian}) are well localized; however, they are not compactly supported. 
One can truncate them with the above mollifier to yield compactly supported functions. 
Consider the normalized Gaussian function
\begin{equation}
    \psi(x)=\pi^{-\frac{1}{4}}\mathrm{e}^{-\frac{x^2}{2}},
\end{equation}
which satisfies $\sigma_x=\sigma_k=1/\sqrt{2}$ and hence $\sigma_x\sigma_k=1/2$. 
It can be truncated with the mollifier as follows:
\begin{equation}
    \varphi(x)
    =
    \dfrac{1}{\|\psi(x)M(x,L,\lambda)\|_2}
    \psi(x)M(x,L,\lambda),
    \label{equ:Bourgain_initial}
\end{equation}
where $\|\cdots\|_2$ refers to the \(L^2(\mathbb{R})\) norm. 
One such function is shown in Fig.~\ref{fig:Bourgain_mollifier}(b). 

As $L$ grows larger, the smooth and compactly supported 
function $\varphi(x)$ approaches the Gaussian wave packet, so that it 
achieves near-minimal uncertainties,
\begin{equation}
    \sigma_x(\varphi)<\dfrac{1}{\sqrt{2}}+\varepsilon,
\end{equation}
and 
\begin{equation}
    \sigma_k(\varphi)<\dfrac{1}{\sqrt{2}}+\varepsilon,
\end{equation}
where $\varepsilon$ is a small positive number. 
This means that both the function $\varphi(x)$ 
and its Fourier transform $\tilde{\varphi}(k)$ are well localized.   
Such truncated Gaussian functions will be used as the main components to construct well localized basis functions
in phase space.  

A set of functions \(\mathcal{F}\) is dense in \(L^2(\mathbb{R})\) 
if for every function \(g \in L^2\) and any \(\epsilon > 0\),
there exists a function \(f \in \mathcal{F}\) such that \(\|g-f\|_2<\epsilon\).
    
Here are two examples of dense sequences in \(L^2\).
The first example consists of polynomials with rational coefficients truncated by the mollifier function:
\begin{equation}
\label{eq:dense1}
    f_j(x)
    =
    \sum_{\ell=0}^{n_j}
    \left(a_{\ell}^{(j)}+ib_{\ell}^{(j)}\right)
    x^\ell M(x,L_j,\lambda_j),
\end{equation}
where $n_j\in\mathbb{N}$ and $a_{\ell}^{(j)},b_{\ell}^{(j)},L_j,\lambda_j\in\mathbb{Q}$.    
The second example is a set of trigonometric polynomials with rational coefficients truncated by the mollifier function:
\begin{equation}
    f_j(x)
    =
    \sum_{\ell=-n_j}^{n_j}
    \left(a_{\ell}^{(j)}+ib_{\ell}^{(j)}\right)
    \mathrm{e}^{i\ell x}M(x,L_j,\lambda_j).
\end{equation}
As these coefficients are rational numbers, these two dense sequences are countable. 
Therefore, we index the sequences as \(\{f_j\}\).

The dense sequence is selected before the construction procedure starts, and it remains unchanged 
during the whole construction process.
The dense sequence is linearly combined with the orthonormal basis functions we are going to 
build in order to ensure the completeness of the final basis. We will elaborate on it in the following text.
Also, we will change some notations in Bourgain's original paper to make them more consistent with our paper.

\subsection{Basis for a finite lattice in phase space}

Consider a finite region in phase space, where $x\in [X_0, X_0+Th]$ and $k\in [0, Th]$, with $T$ being 
a large positive integer and $h>0$. We divide this region into a $T\times T$ lattice, with $h$ being the lattice constant
along both the $x$ and $k$ directions. The size of each cell is therefore $h^2$. Since 
it is not necessary that $h^2=2\pi$, these cells are not necessarily the Planck cells discussed before. 
We label these cells with two integer indices $\alpha,\beta=0,1,2,\cdots,T-1$. For each cell, we assign the 
following function:
\begin{equation}
    r_{\alpha,\beta}(x)
    =
    \varphi\left(x-X_0-(\alpha+1/2)h\right)
    \exp(i\beta h x),
\end{equation}
where $\varphi(x)$ is the function defined in Eq.~(\ref{equ:Bourgain_initial}) with $L=h$, so that 
it is compactly supported in $[-h/2,h/2]$.
Such defined $r_{\alpha,\beta}(x)$ is centered at 
\begin{equation}
    \left(X_0+(\alpha+1/2)h,\beta h\right)
\end{equation}
in phase space.
When $h$ is large enough, $\varphi(x)$ is close to the Gaussian function $\psi(x)$ with minimum uncertainty. 
    
We denote this family of functions as 
\(\mathcal{R}=\{r_{\alpha,\beta}(x)\}\), which forms a periodic tiling in the finite region of phase space. 
It is evident that $r_{\alpha,\beta}$ and $r_{\alpha',\beta'}$ 
are disjointly supported when $\alpha\neq\alpha'$. For $\alpha=\alpha'$, the overlap is
\begin{equation}
    \Braket{r_{\alpha,\beta}|r_{\alpha',\beta'}}
    =
    \delta_{\alpha\alpha'}
    e^{i(\beta'-\beta)hX_\alpha}
    \int_{-\infty}^{\infty}\mathrm{d}u\,
    |\varphi(u)|^2 e^{i(\beta'-\beta)hu},
\end{equation}
where
\begin{equation}
    X_\alpha=X_0+(\alpha+1/2)h.
\end{equation}
Thus, the magnitude of the overlap is controlled by the Fourier transform of $|\varphi|^2$ at frequency $(\beta'-\beta)h$. 
When $|\beta'-\beta|h$ becomes larger, the overlap decays rapidly due to the smooth compact support of $\varphi(x)$.
    
One can employ various orthogonalization methods, such as Gram--Schmidt orthogonalization or L\"owdin orthogonalization \cite{lowdin1950non}, to orthogonalize $\mathcal{R}$.
Bourgain chose to use Gram--Schmidt orthogonalization in his original paper \cite{bourgain1988remark}.
Here we provide an alternative orthogonalization method, namely L\"owdin orthogonalization, owing to its superior numerical stability.
    
Since the functions in $\mathcal{R}$ with different $\alpha$ are already disjointly supported,
we only need to orthogonalize the functions with the same $\alpha$ but different $\beta$.
We denote the overlap matrix of $\{r_{\alpha,\beta}\}$ with different $\beta$ as $\mathfrak S$, whose elements are
\begin{equation}
   \mathfrak S_{\beta,\beta'}
   =
   \Braket{r_{\alpha,\beta}|r_{\alpha,\beta'}}.
\end{equation}
Then the orthonormal functions $\{s_{\alpha,\beta}\}$ can be obtained by
\begin{equation}
    s_{\alpha,\beta}
    =
    \sum_{\beta'=0}^{T-1}
    r_{\alpha,\beta'}
    \left(\mathfrak S\right)^{-\frac{1}{2}}_{\beta',\beta}.
\end{equation}
In the above computation, one needs to fix $\alpha$. After that, one can either repeat the computation 
for different $\alpha$ or move the functions $s_{\alpha,\beta}$ to lattices with different $\alpha$. 
In the end, we obtain a set of functions 
\begin{equation}
    \mathcal{S}=\{s_{\alpha,\beta}(x)\}
\end{equation}
that are orthonormal to each other. 

It can be inferred that, as the overlap $\Braket{r_{\alpha,\beta}|r_{\alpha',\beta'}}$ decreases rapidly 
with the increase of $h$, for sufficiently large $h$, the overlap is already small even before the orthogonalization process. 
Thus, orthogonalization introduces only a marginal perturbation to the functions. Therefore, 
for sufficiently large $h$, the functions in $\mathcal{S}$ should be well localized,
\begin{equation}
    \sigma_x(s)<\dfrac{1}{\sqrt{2}}+2\varepsilon,
\end{equation}
and 
\begin{equation}
    \sigma_k(s)<\dfrac{1}{\sqrt{2}}+2\varepsilon.
\end{equation}
Bourgain proved the above statement for the Gram--Schmidt orthogonalization method in the original paper. 
We can extend his proof for the L\"owdin orthogonalization method (see Appendix \ref{sec:Appendix_D}).

\subsection{Mixing with the dense sequence}

In the above, we have just built an orthonormal system $\mathcal{S}$ with $T\times T$ smooth 
compactly supported functions, each of which is arbitrarily close 
to the Gaussian wave packet with minimum uncertainty. The problem is that this system 
is not complete even for the finite region in phase space covered by the lattice. 
As a remedy, we need to use the following technique. The effect of this technique is not obvious 
and can be appreciated after the whole construction procedure is presented.  

We choose one of the functions in the dense sequence in Eq.~(\ref{eq:dense1}):
\begin{equation}
\label{eq:dense11}
    f(x)
    =
    \sum_{\ell=0}^{n}
    \left(a_{\ell}+ib_{\ell}\right)x^\ell M(x,L_\ell,\lambda_\ell).
\end{equation}
This function $f(x)$ is chosen so that it has disjoint support with all the functions in $\mathcal{S}$. 
The strategy is to construct the orthonormal basis 
\begin{equation}
    \mathcal{B}=\{b_{\alpha,\beta}\}
\end{equation}
by mixing this function with $\mathcal{S}$ with a small weight. 

To make the recursive orthogonalization transparent, we enumerate the two-index family 
$\mathcal{S}=\{s_{\alpha,\beta}\}$ by a single index:
\begin{equation}
    l=\alpha T+\beta,
    \qquad
    s_l\equiv s_{\alpha,\beta},
    \qquad
    b_l\equiv b_{\alpha,\beta},
\end{equation}
where $l=0,1,\ldots,T^2-1$. 
Following Bourgain's construction, we define
\begin{equation}
    b_l
    =
    \dfrac{\mu}{T}f
    +
    \sum_{r=0}^{l-1}\sigma_r s_r
    +
    \gamma_l s_l,
    \label{eq:Bourgain_bl}
\end{equation}
where $\mu$, $\sigma_r$, and $\gamma_l$ are real parameters to be determined by the orthonormality requirements. 
Here $\mu/T$ is the small mixing weight of the dense-sequence component $f$, while $\gamma_l$ is the principal coefficient of the localized function $s_l$.

Since $f$ has disjoint support with all functions in $\mathcal{S}$, we have 
$\braket{f|s_l}=0$ for all $l$. Moreover, the functions $\{s_l\}$ are already orthonormal. 
Therefore, for $0\leq p<l\leq T^2-1$, the orthogonality condition 
$\braket{b_p|b_l}=0$ gives
\begin{equation}
    \dfrac{\mu^2}{T^2}\|f\|_2^2
    +
    \sum_{r=0}^{p-1}\sigma_r^2
    +
    \gamma_p\sigma_p
    =
    0.
    \label{eq:Bourgain_orthogonality}
\end{equation}
Here the term $\gamma_p\sigma_p$ arises because the coefficient of $s_p$ in $b_p$ is $\gamma_p$, whereas the coefficient of $s_p$ in $b_l$ is $\sigma_p$.

The normalization condition $\|b_l\|_2=1$ gives
\begin{equation}
    \dfrac{\mu^2}{T^2}\|f\|_2^2
    +
    \sum_{r=0}^{l-1}\sigma_r^2
    +
    \gamma_l^2
    =
    1.
    \label{eq:Bourgain_normalization}
\end{equation}
Equations~(\ref{eq:Bourgain_orthogonality}) and~(\ref{eq:Bourgain_normalization}) determine the coefficients recursively. 
For $\mu<1/2$, these equations imply the estimates
\begin{equation}
    |\sigma_l|
    \leq
    \dfrac{\mu}{T^2},
\end{equation}
and
\begin{equation}
    1-\gamma_l
    <
    2\dfrac{\mu}{T},
\end{equation}
for any $l=0,1,\ldots,T^2-1$. 
Consequently,
\begin{equation}
    \|b_l-s_l\|_2
    \leq
    3\dfrac{\mu}{T}.
\end{equation}
Equivalently, in the original two-index notation,
\begin{equation}
    \|b_{\alpha,\beta}-s_{\alpha,\beta}\|_2
    \leq
    3\dfrac{\mu}{T},
\end{equation}
for any $\alpha,\beta=0,1,\ldots,T-1$.

For sufficiently large $T$, this small perturbation does not destroy the localization inherited from $\mathcal{S}$. 
Thus, the resulting functions $b_{\alpha,\beta}$ satisfy
\begin{equation}
    \sigma_x(b_{\alpha,\beta})<\dfrac{1}{\sqrt{2}}+3\varepsilon,
\end{equation}
and 
\begin{equation}
    \sigma_k(b_{\alpha,\beta})<\dfrac{1}{\sqrt{2}}+3\varepsilon.
\end{equation}
This shows that the localization of the functions in $\mathcal{S}$ is only affected marginally by
the addition of the dense-sequence component $f(x)$. 

\begin{figure}[htbp]
    \centering
    \includegraphics[width=\columnwidth]{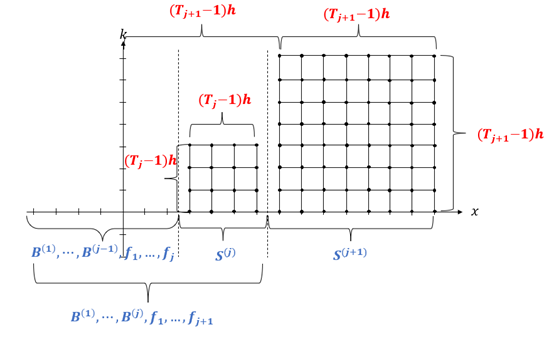}
    \caption{Schematic diagram of the $j$-th and $(j+1)$-th steps in the construction of Bourgain's basis in phase space.}
    \label{fig:Bourgain_basis} 
\end{figure}

\subsection{Construction of Bourgain's basis}

We are now ready to present 
Bourgain's method for constructing the nonperiodic basis. His procedure has 
infinitely many steps, and we describe it inductively. 

We assume that we have constructed a series of orthonormal families, 
$\{\mathcal{B}^{(1)},\ldots,\mathcal{B}^{(j)}\}$. As shown schematically in Fig.~\ref{fig:Bourgain_basis},
at each step the finite lattice block is placed sufficiently far along the $x$ direction so that its support 
is disjoint from the previously constructed functions and from the compact support of the dense function used at that step.
 
The following are the steps to construct the next basis set $\mathcal{B}^{(j+1)}$:
\begin{enumerate}
    \item We set up a new lattice block $T_{j+1}\times T_{j+1}$ sufficiently far from the previously constructed blocks. 
    As described above, we construct the orthonormal family $\mathcal{S}^{(j+1)}$ on this block. 
 
    \item We take the next function $f_{j+1}(x)$ in the fixed dense sequence. The new block is chosen sufficiently far away 
    so that $\{\mathcal{B}^{(1)},\ldots,\mathcal{B}^{(j)}\}$ and $f_{j+1}$ are contained within a region disjoint from the support of $\mathcal{S}^{(j+1)}$. 
    Let us define a projected function 
    \begin{equation}
        \tilde{f}_{j+1}(x)
        =
        f_{j+1}
        -
        P_{[\mathcal{B}^{(1)},\ldots,\mathcal{B}^{(j)}]}f_{j+1},
    \end{equation} 
    where $P_{[\mathcal{B}^{(1)},\ldots,\mathcal{B}^{(j)}]}$ is the projection onto the subspace spanned by 
    the basis functions in $\{\mathcal{B}^{(1)},\ldots,\mathcal{B}^{(j)}\}$. Therefore, $\tilde{f}_{j+1}(x)$ is orthogonal to 
    all the basis functions constructed in the first $j$ iterations.  
 
    \item In the final step of this iteration, we construct the basis set $\mathcal{B}^{(j+1)}$ by combining 
    $\mathcal{S}^{(j+1)}$ and $\tilde{f}_{j+1}(x)$ as described above. 
\end{enumerate}
We repeat this iteration infinitely many times and obtain a sequence of basis sets $\mathcal{B}^{(j)}$.  

To prove completeness, it is sufficient to show that no element of the fixed dense sequence is orthogonal to the 
closed span of the constructed system. More quantitatively, Bourgain's construction ensures that, for any function $f_j(x)$ in the dense sequence,
\begin{equation}
    \|P_{[\mathcal{B}^{(1)},\ldots,\mathcal{B}^{(j)}]}f_j\|_2> v(\varepsilon),
\end{equation}
where $v(\varepsilon)>0$ is a function of $\varepsilon$. This condition can indeed be verified. 
 
The norm of the projection $\|P_{[\mathcal{B}^{(1)},\ldots,\mathcal{B}^{(j)}]}f_j\|_2$ can be decomposed as follows:
\begin{equation}
    \|P_{[\mathcal{B}^{(1)},\ldots,\mathcal{B}^{(j)}]}f_j\|_2^2
    =
    \|P_{[\mathcal{B}^{(1)},\ldots,\mathcal{B}^{(j-1)}]}f_j\|_2^2
    +
    \|P_{[\mathcal{B}^{(j)}]}\tilde{f}_j\|_2^2,
\end{equation}
where $P_{[\mathcal{B}^{(j)}]}\tilde{f}_j$ is the orthogonal projection of $\tilde{f}_j$ 
onto the space spanned by $\mathcal{B}^{(j)}$.
Here we have invoked the property that $\tilde{f}_j$ is orthogonal to $\{\mathcal{B}^{(1)},\ldots,\mathcal{B}^{(j-1)}\}$.
Let
\begin{equation}
    \|P_{[\mathcal{B}^{(1)},\ldots,\mathcal{B}^{(j-1)}]}f_j\|_2 = a,
\end{equation}
where $a\in[0,1]$ assuming that $f_j$ is normalized. Therefore,
\begin{equation}
    \|\tilde{f}_j\|_2 = \sqrt{1-a^2}. 
\end{equation}
We can estimate $\|P_{[\mathcal{B}^{(j)}]}\tilde{f}_j\|_2$ as follows:
\begin{equation}
    \begin{aligned}
        \|P_{[\mathcal{B}^{(j)}]}\tilde{f}_j\|_2^2
        &=
        \sum_{\alpha,\beta}
        |\Braket{b^{(j)}_{\alpha,\beta}|\tilde{f}_j}|^2\\
        &\geq
        \left(\dfrac{\mu}{T_j}\right)^2 T_j^2 \|\tilde{f}_j\|_2^4 \\
        &=
        \mu^2 \|\tilde{f}_j\|_2^4 .
    \end{aligned}
\end{equation}
Consequently,
\begin{equation}
    \begin{aligned}
        v(\varepsilon)
        &=
        \sqrt{
        \|P_{[\mathcal{B}^{(1)},\ldots,\mathcal{B}^{(j-1)}]}f_j\|_2^2
        +
        \|P_{[\mathcal{B}^{(j)}]}\tilde{f}_j\|_2^2
        }\\
        &\geq
        \sqrt{a^2+(1-a^2)^2\mu^2}\\
        &\geq
        \dfrac{\mu}{2}>0.
    \end{aligned}
\end{equation}
This proves the completeness of Bourgain's basis. 
 
\begin{figure}[htbp]
    \centering
    \includegraphics[width=\columnwidth]{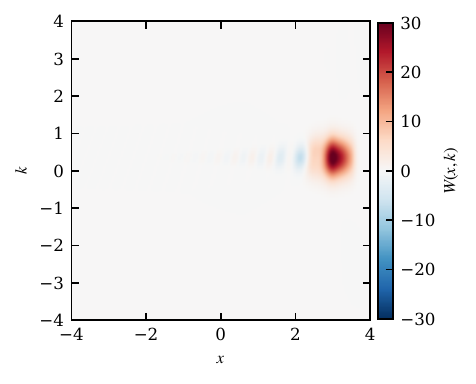}
    \caption{Schematic Wigner-type illustration of a Bourgain-like basis element. 
The dominant Gaussian-like core near $(x,k)\approx(3,1/3)$ represents the localized component inherited from a 
finite orthonormal lattice block, while the weak oscillatory tail extending toward smaller $x$ 
mimics the perturbative dense-sequence component introduced in Bourgain's construction. 
The alternating red and blue fringes represent Wigner-type interference between the localized core and the perturbative component. 
This figure is schematic and is intended only to visualize the qualitative structure of a Bourgain-like basis 
element, rather than to represent an exact numerical implementation of Bourgain's original construction.}
    \label{fig:wigner_bourgain}
\end{figure}

Bourgain's construction proves that a complete orthonormal basis in phase space can asymptotically approach the lower bound of the uncertainty principle. 
However, his construction is purely existential in nature and has almost no computational feasibility. In practical calculations, 
the truncated basis and Wannier basis presented earlier are more computationally tractable.
\section{Conclusion}
\label{sec:6}

In this review, we have summarized the main ideas and recent developments in quantum phase-space theory. 
Starting from classical phase space, we discussed how the phase-space viewpoint provides an intuitive description of both 
regular and chaotic dynamics, while also emphasizing why a direct extension to quantum mechanics is obstructed by the uncertainty principle.

We first reviewed three standard quasi-probability distributions: the Wigner function, the Husimi $Q$ function, and the Glauber--Sudarshan $P$ function. 
These representations provide powerful ways to visualize quantum states in phase space, but each has intrinsic limitations. 
The Wigner function can take negative values, the Husimi $Q$ function is smoothed by coherent-state coarse graining and 
does not reproduce exact marginal distributions, and the Glauber--Sudarshan $P$ function may become highly singular for nonclassical pure states. 
These features show that quasi-probability distributions, although useful, do not directly provide a genuine probability distribution over mutually exclusive Planck cells.

We then discussed von Neumann's alternative viewpoint: constructing an orthonormal basis localized in Planck cells. 
The truncated basis gives a simple and intuitive realization of this idea, but its sharp cutoff in position space leads to poor localization in momentum space. 
The quantum Wannier basis, constructed through Kohn and L\"owdin orthogonalization, provides a more refined realization. 
It yields a complete orthonormal basis and therefore defines a unitary mapping from Hilbert space to a discretized quantum phase space. 
Consequently, the expansion coefficients in this basis give a genuine probability distribution over Planck cells.

The applications discussed in this review illustrate the usefulness of this basis-based formulation. 
For the quantum kicked rotor, the truncated Planck-cell basis provides a natural way to construct 
quantum Poincar\'e sections and to visualize signatures of classical phase-space structures. 
For finite-dimensional systems such as the two-site Bose--Hubbard model, an analogous 
number-phase Planck-cell basis can be constructed in the Fock representation. 
These examples show that von Neumann-type quantum phase spaces offer a flexible 
framework for connecting quantum states with classical or semiclassical phase-space intuition.

A central mathematical limitation of translationally symmetric phase-space bases is the Balian--Low theorem. 
Although the quantum Wannier basis is well localized for practical finite cutoffs, exact 
translational symmetry forces the simultaneous second moments in the infinite limit to diverge. 
Bourgain's nonperiodic construction shows that this obstruction can be avoided if strict translational symmetry is relaxed. 
It provides an existence proof of a complete orthonormal basis whose localization can 
approach the Gaussian uncertainty limit arbitrarily closely. 
Although this construction is not directly practical for numerical computation, 
it serves as an important conceptual benchmark for understanding the ultimate localization limits of orthonormal phase-space bases.

Overall, quantum phase-space theory contains two complementary directions. 
Quasi-probability distributions provide continuous phase-space pictures and are especially 
useful for visualizing interference and semiclassical behavior. 
Planck-cell bases, in contrast, provide unitary maps from Hilbert space to discrete 
phase space and hence yield genuine probability distributions. 
Further development of quantum phase-space bases, especially more computationally 
tractable constructions with improved localization, may be useful for quantum dynamics, quantum simulation, and semiclassical analysis.
\appendix\section{Orthonormality of the Spatial-Translation-Symmetric Wannier Basis}
\label{sec:Appendix_A}

In this appendix, we prove the orthonormality of the spatial-translation-symmetric
Wannier basis $\{\ket{w_{j_x,j_k}}\}$ constructed in Section \ref{sec:wannierb}.
We use the same symmetric Fourier convention as in Section \ref{sec:wannierb}.

After L\"owdin orthogonalization, the momentum-space functions $\tilde{w}_{j_k}(k)$
satisfy
\begin{equation}
    \sum_{n \in \mathbb{Z}}
    \tilde{w}_{j_k}^{*}(k+n k_0)
    \tilde{w}_{j_k'}(k+n k_0)
    =
    \frac{1}{k_0}\delta_{j_k,j_k'},
    \label{eq:Appendix_A_lowdin_identity}
\end{equation}
where $k\in[0,k_0)$ and $k_0=2\pi/x_0$. The factor $1/k_0$ follows from the
normalization convention
$\tilde w_{j_k}(k+nk_0)=u_{k,j_k}(n)/\sqrt{k_0}$ used in Section \ref{sec:wannierb}.
The spatial translation
\begin{equation}
    w_{j_x,j_k}(x)=w_{j_k}(x-j_xx_0)
\end{equation}
gives
\begin{equation}
    \tilde{w}_{j_x,j_k}(k)
    =
    e^{-ikj_xx_0}\tilde{w}_{j_k}(k).
    \label{eq:Appendix_A_shift}
\end{equation}

Let $\Delta j_x=j_x-j_x'$. Using Eq.~\eqref{eq:Appendix_A_shift}, we obtain
\begin{equation}
\begin{aligned}
\Braket{w_{j_x,j_k}|w_{j_x',j_k'}}
&=
\int_{-\infty}^{+\infty}
\tilde{w}_{j_k}^{*}(k)
\tilde{w}_{j_k'}(k)
e^{ikx_0\Delta j_x}
\,\mathrm{d}k                                                        \\
&=
\int_{0}^{k_0}
e^{ikx_0\Delta j_x}
\left[
\sum_{n\in\mathbb{Z}}
\tilde{w}_{j_k}^{*}(k+n k_0)
\tilde{w}_{j_k'}(k+n k_0)
\right]
\,\mathrm{d}k                                                        \\
&=
\frac{\delta_{j_k,j_k'}}{k_0}
\int_{0}^{k_0}
e^{ikx_0\Delta j_x}
\,\mathrm{d}k                                                        \\
&=
\delta_{j_x,j_x'}\delta_{j_k,j_k'} .
\end{aligned}
\label{eq:Appendix_A_orthonormality}
\end{equation}
In the second equality, each interval $[nk_0,(n+1)k_0)$ has been shifted back to
$[0,k_0)$. The additional phase factor is unchanged because
\begin{equation}
    e^{in k_0x_0\Delta j_x}
    =
    e^{i2\pi n\Delta j_x}
    =
    1,
\end{equation}
where $n$ and $\Delta j_x$ are integers. The last equality follows from
\begin{equation}
    \frac{1}{k_0}
    \int_0^{k_0}
    e^{ikx_0(j_x-j_x')}
    \,\mathrm{d}k
    =
    \delta_{j_x,j_x'} .
\end{equation}
Therefore, the spatial-translation-symmetric Wannier basis satisfies
\begin{equation}
    \Braket{w_{j_x,j_k}|w_{j_x',j_k'}}
    =
    \delta_{j_x,j_x'}\delta_{j_k,j_k'} .
\end{equation}

\section{Balian--Low Constraint on the Infinite-Cutoff Wannier Basis}
\label{sec:Appendix_B}

In this appendix, we explain why the uncertainty product of the
spatial-translation-symmetric Wannier basis must diverge in the limit
$J_k\to\infty$. Since spatial translations do not change the variances
in $x$ and $k$, it is sufficient to consider the family
$\{w_{0,j_k}^{(J_k)}\}$ and define
\begin{equation}
    \Sigma(J_k)
    =
    \max_{|j_k|\leq J_k}
    \left[
    \sigma_x\!\left(w_{0,j_k}^{(J_k)}\right)
    \sigma_k\!\left(w_{0,j_k}^{(J_k)}\right)
    \right].
    \label{eq:Appendix_B_Sigma}
\end{equation}
We show that $\Sigma(J_k)$ cannot remain bounded as $J_k\to\infty$.

For fixed reduced momentum $k\in[0,k_0)$, define
\begin{equation}
    f_{k,j}(n)
    =
    \tilde g_j(k+n k_0),
    \qquad n\in\mathbb Z .
\end{equation}
The L\"owdin overlap matrix for the momentum-cell indices is
\begin{equation}
    \mathsf S_{ij}^{(J_k)}(k)
    =
    \sum_{n\in\mathbb Z}
    \tilde g_i^{*}(k+n k_0)
    \tilde g_j(k+n k_0),
    \qquad |i|,|j|\leq J_k .
    \label{eq:Appendix_B_overlap}
\end{equation}
For the Gabor-type seed functions
\begin{equation}
    g_{j_x,j_k}(x)
    =
    g(x-j_xx_0)\exp(ij_k k_0x),
    \label{eq:Appendix_B_seed_gabor}
\end{equation}
one has
\begin{equation}
    \tilde g_j(k+n k_0)
    =
    \tilde g_0\!\left(k+(n-j)k_0\right).
\end{equation}
Hence, in the infinite-cutoff limit,
\begin{equation}
\begin{aligned}
    \mathsf S_{i+a,j+a}^{(\infty)}(k)
    &=
    \sum_{n\in\mathbb Z}
    \tilde g_{i+a}^{*}(k+n k_0)
    \tilde g_{j+a}(k+n k_0)                                  \\
    &=
    \sum_{m\in\mathbb Z}
    \tilde g_i^{*}(k+m k_0)
    \tilde g_j(k+m k_0)                                      \\
    &=
    \mathsf S_{ij}^{(\infty)}(k),
\end{aligned}
\label{eq:Appendix_B_shift_invariance}
\end{equation}
where $m=n-a$ has been used. Therefore
\begin{equation}
    \mathsf S_{ij}^{(\infty)}(k)
    =
    s_{i-j}(k),
    \label{eq:Appendix_B_toeplitz}
\end{equation}
i.e., the infinite overlap matrix is a Toeplitz, or convolution, operator
in the $j_k$ index.

The decay away from the diagonal follows from the localization of the seed
function in momentum space. For example, with the Gaussian seed used in
Section \ref{sec:wannierb}, one may write
$\tilde g_j(k)=\exp[-(k/k_0-j)^2]$ up to normalization. Setting
$q=k/k_0\in[0,1)$, Eq.~\eqref{eq:Appendix_B_overlap} gives
\begin{equation}
\begin{aligned}
    |\mathsf S_{ij}^{(\infty)}(k)|
    &=
    \sum_{n\in\mathbb Z}
    e^{-(q+n-i)^2-(q+n-j)^2}                                  \\
    &=
    e^{-(i-j)^2/2}
    \sum_{n\in\mathbb Z}
    e^{-2[q+n-(i+j)/2]^2}                                     \\
    &\leq
    C_0 e^{-(i-j)^2/2},
\end{aligned}
\label{eq:Appendix_B_gaussian_decay}
\end{equation}
where
\begin{equation}
    C_0=\sup_{t\in\mathbb R}\sum_{n\in\mathbb Z}e^{-2(n+t)^2}\approx 1.2713 .
\end{equation}
Thus the overlap matrix is not only Toeplitz in the infinite limit, but also
exponentially localized near its diagonal.

For finite $J_k$, Eq.~\eqref{eq:Appendix_B_overlap} is the restriction of this
infinite Toeplitz matrix to the interval $-J_k\leq j\leq J_k$. The only
violation of translation covariance comes from the two boundaries. If the
row index $i$ is at distance
\begin{equation}
    r_i=J_k-|i|
\end{equation}
from the nearest boundary, then the omitted tail is bounded by
\begin{equation}
    \sum_{|j|>J_k}
    |\mathsf S_{ij}^{(\infty)}(k)|
    \leq
    C_0
    \sum_{|m|>r_i}
    e^{-m^2/2},
    \label{eq:Appendix_B_boundary_tail}
\end{equation}
which is exponentially small when $r_i\gg1$. Therefore the finite matrix
recovers the translation-covariant infinite matrix away from the boundaries,
and the boundary region moves to infinity as $J_k\to\infty$.

The L\"owdin orthogonalization is
\begin{equation}
    u_{k,j}^{(J_k)}
    =
    \sum_{i=-J_k}^{J_k}
    f_{k,i}
    \left[\mathsf S^{(J_k)}(k)^{-1/2}\right]_{ij}.
    \label{eq:Appendix_B_lowdin}
\end{equation}
In the infinite limit, $\mathsf S^{(\infty)}(k)$ commutes with the shift operator
in the $j_k$ index. Hence any function of it, in particular
$\mathsf S^{(\infty)}(k)^{-1/2}$, also commutes with this shift operator. Its
matrix elements therefore depend only on the index difference:
\begin{equation}
    \left[\mathsf S^{(\infty)}(k)^{-1/2}\right]_{ij}
    =
    h_{i-j}(k).
    \label{eq:Appendix_B_inverse_toeplitz}
\end{equation}
Consequently, the L\"owdin-orthogonalized vectors satisfy the same translation
covariance in the momentum-cell index. In $k$ space this gives
\begin{equation}
    \tilde w_{j_k}(k)
    =
    \tilde w_0(k-j_k k_0),
    \label{eq:Appendix_B_k_translation}
\end{equation}
or equivalently in real space,
\begin{equation}
    w_{j_k}(x)
    =
    e^{ij_k k_0x}w_0(x).
    \label{eq:Appendix_B_real_modulation}
\end{equation}
Combining this $k$-direction covariance with the spatial translation
$w_{j_x,j_k}(x)=w_{j_k}(x-j_xx_0)$, the infinite-cutoff basis takes the
Gabor form
\begin{equation}
    w_{j_x,j_k}(x)
    =
    e^{ij_k k_0x}w(x-j_xx_0),
    \qquad x_0k_0=2\pi .
    \label{eq:Appendix_B_gabor_form}
\end{equation}

The Balian--Low theorem states that a complete orthonormal Gabor basis at
critical density cannot be generated by a window function with finite second
moments in both conjugate variables. In the present notation, this gives
\begin{equation}
    \sigma_x(w)\sigma_k(w)=\infty .
    \label{eq:Appendix_B_balian_low}
\end{equation}
If $\Sigma(J_k)$ remained bounded as $J_k\to\infty$, the limiting construction
would give a complete orthonormal Gabor basis with finite uncertainty product,
contradicting Eq.~\eqref{eq:Appendix_B_balian_low}. Therefore,
\begin{equation}
    \lim_{J_k\to\infty}\Sigma(J_k)=\infty .
    \label{eq:Appendix_B_divergence}
\end{equation}

This explains the unavoidable divergence caused by the Balian--Low theorem.
The argument does not determine the precise growth rate of $\Sigma(J_k)$;
the sub-logarithmic behavior shown in Fig.~\ref{fig:walocal} is a numerical
feature of the present finite-cutoff construction.
\section{Husimi Stroboscopic Portraits of the Quantum Kicked Rotor}
\label{app:husimi-qkr}

In the main text, the quantum Poincar\'e sections of the kicked rotor are constructed
from the truncated Planck-cell basis $\{\ket{X,P}\}$. 
To compare this orthonormal-basis representation with a more conventional
phase-space visualization, we also compute Husimi-type stroboscopic portraits
for the same quantum dynamics.

For the torus representation of the kicked rotor, we use coherent states
$\ket{x,p;\sigma}$ localized near the phase-space point $(x,p)$, with
$(x,p)\in[0,2\pi)\times[0,2\pi)$. 
For a quantum state $\ket{\psi_t}$ immediately after the $t$-th kick, we define
the Husimi intensity as
\begin{equation}
    Q_t(x,p)
    =
    |\Braket{x,p;\sigma|\psi_t}|^2 ,
\end{equation}
where $\sigma$ controls the coherent-state width. 
Here we omit an overall normalization factor, since only the relative
phase-space intensity is used for visualization. 
The stroboscopic Husimi portrait is obtained by averaging this intensity over
successive kicks:
\begin{equation}
    \overline{Q}_T(x,p)
    =
    \frac{1}{T}
    \sum_{t=0}^{T-1}
    Q_t(x,p).
    \label{eq:Appendix_C_time_averaged_Husimi}
\end{equation}

\begin{figure*}[t]
    \centering
    \includegraphics[width=\textwidth]{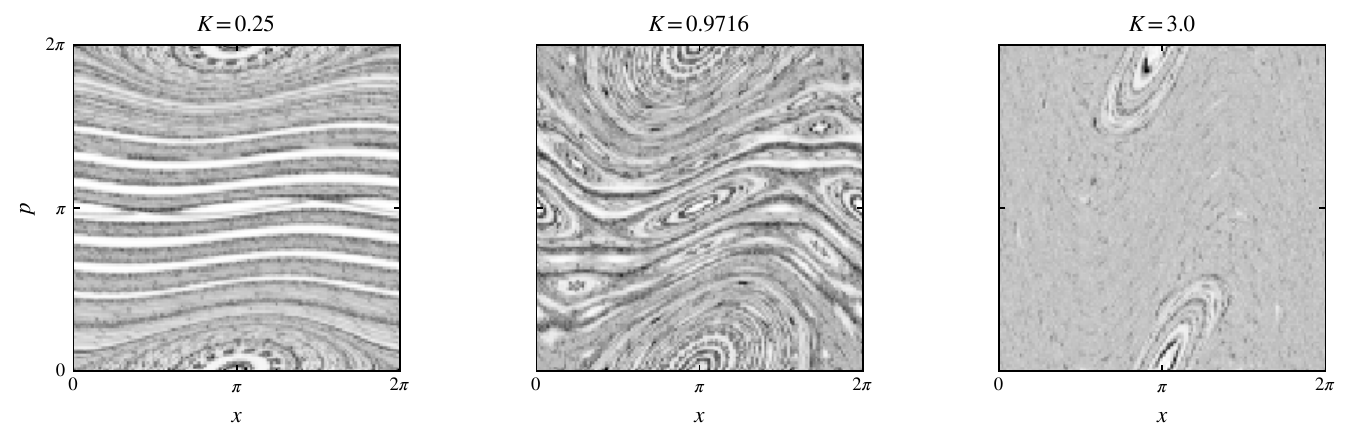}
    \caption{Time-averaged Husimi phase-space portraits of the quantum kicked rotor
    for three kicking strengths, shown from left to right: $K=0.25$, $K=0.9716$,
    and $K=3.0$. The Husimi intensity is evaluated stroboscopically after each
    kick and averaged over $T=60$ kicks, using the same initial state and effective
    Planck constant as in Fig.~\ref{fig:poincare_sections}. The plots visualize
    coherent-state-smoothed phase-space densities and should not be interpreted as
    probabilities over mutually exclusive Planck cells.}
    \label{fig:husimi_qkr_appendix}
\end{figure*}

Figure~\ref{fig:husimi_qkr_appendix} shows the resulting time-averaged Husimi
portraits for the same three kicking strengths as in Fig.~\ref{fig:poincare_sections}. 
The main classical structures remain visible after coherent-state smoothing. 
For $K=0.25$ (left panel), the portrait exhibits regular band-like structures. 
For $K=0.9716$ (middle panel), it shows a mixed phase-space pattern with both
regular and irregular regions. 
For $K=3.0$ (right panel), the distribution is broadly spread over the chaotic
sea, with only small regular islands remaining.

The comparison highlights the distinction between the two phase-space
representations. 
The Husimi representation is positive and visually close to a classical
phase-space density, but it is based on nonorthogonal and overcomplete coherent
states. 
Therefore, $\overline{Q}_T(x,p)$ is a smoothed phase-space intensity rather than
a probability distribution over an orthonormal set of Planck cells. 
In contrast, the truncated basis $\{\ket{X,P}\}$ used in the main text is
orthonormal, so the coefficients
\begin{equation}
    P_{X,P}(t)
    =
    |\Braket{X,P|\psi_t}|^2
\end{equation}
define genuine probabilities on von Neumann's discrete quantum phase space. 
This is the main reason why the Planck-cell construction gives a more direct
quantum analog of a coarse-grained classical Poincar\'e section.
\section{Localization Stability under L\"owdin Orthogonalization}
\label{sec:Appendix_D}

In this appendix, we justify that the L\"owdin orthogonalization used in
Section \ref{sec:5} does not substantially change the localization of the finite
Bourgain lattice functions $r_{\alpha,\beta}^{(j)}$. 
We focus on the $j$-th step and fix the spatial index $\alpha$. 
Since functions with different $\alpha$ have disjoint supports, only the functions
with the same $\alpha$ and different $\beta$ need to be orthogonalized.

Let
\begin{equation}
    \mathcal R_{\alpha}^{(j)}
    =
    \{r_{\alpha,\beta}^{(j)}\}_{\beta=0}^{T_j-1}.
\end{equation}
For $\beta\neq\beta'$, define $m=\beta'-\beta$. 
Up to an irrelevant phase factor coming from the center of the spatial cell, the overlap is
\begin{equation}
    \Braket{r_{\alpha,\beta}^{(j)}|r_{\alpha,\beta'}^{(j)}}
    =
    \int_{-\infty}^{+\infty}
    e^{imhx}|\varphi(x)|^2\,\mathrm{d}x .
    \label{eq:Appendix_D_overlap}
\end{equation}
Equivalently, since $\varphi(x)$ is real and even in the present construction, this is the cosine overlap used in Section \ref{sec:5}.

Recall that
\begin{equation}
    \varphi(x)
    =
    \frac{\psi(x)M(x,h,\lambda)}
    {\|\psi M\|_2},
    \qquad
    \psi(x)=\pi^{-1/4}e^{-x^2/2},
\end{equation}
where $M(x,h,\lambda)=1$ for $|x|\leq \lambda h/2$ and
$M(x,h,\lambda)=0$ for $|x|\geq h/2$. Let
\begin{equation}
    N_h=\|\psi M(\cdot,h,\lambda)\|_2 .
\end{equation}
Using
\begin{equation}
    \int_{-\infty}^{+\infty}
    e^{-x^2}e^{imhx}\,\mathrm{d}x
    =
    \sqrt{\pi}\,e^{-m^2h^2/4}
\end{equation}
and the Gaussian tail estimate
\begin{equation}
    \int_a^{+\infty}e^{-x^2}\,\mathrm{d}x
    <
    \frac{1}{2a}e^{-a^2},
    \qquad a>0,
\end{equation}
we obtain
\begin{equation}
\begin{aligned}
    \left|
    \Braket{r_{\alpha,\beta}^{(j)}
    |r_{\alpha,\beta'}^{(j)}}
    \right|
    &\leq
    \frac{1}{N_h^2}
    e^{-m^2h^2/4}                                      \\
    &\quad+
    \frac{4}{\sqrt{\pi}N_h^2\lambda h}
    e^{-\lambda^2h^2/4}.
\end{aligned}
\label{eq:Appendix_D_overlap_bound_m}
\end{equation}
In particular, for all $\beta\neq\beta'$,
\begin{equation}
    \left|
    \Braket{r_{\alpha,\beta}^{(j)}
    |r_{\alpha,\beta'}^{(j)}}
    \right|
    \leq
    \delta_j,
    \label{eq:Appendix_D_delta_bound}
\end{equation}
where
\begin{equation}
    \delta_j
    =
    \frac{1}{N_h^2}e^{-h^2/4}
    +
    \frac{4}{\sqrt{\pi}N_h^2\lambda h}
    e^{-\lambda^2h^2/4}.
    \label{eq:Appendix_D_delta}
\end{equation}
Thus the off-diagonal overlaps are exponentially small for large $h$.

Let $\mathfrak S^{(j)}$ be the overlap matrix of
$\mathcal R_{\alpha}^{(j)}$:
\begin{equation}
    \mathfrak S_{\beta,\beta'}^{(j)}
    =
    \Braket{r_{\alpha,\beta}^{(j)}
    |r_{\alpha,\beta'}^{(j)}} .
    \label{eq:Appendix_D_overlap_matrix}
\end{equation}
The diagonal elements are equal to one, while the off-diagonal elements are
bounded by $\delta_j$. By Gershgorin's circle theorem \cite{gershgorin1931uber}, every eigenvalue
$\Lambda$ of $\mathfrak S^{(j)}$ satisfies
\begin{equation}
    1-\rho_j
    \leq
    \Lambda
    \leq
    1+\rho_j,
    \qquad
    \rho_j=(T_j-1)\delta_j .
    \label{eq:Appendix_D_gershgorin}
\end{equation}
For sufficiently large $h$, one can make $\rho_j<1$, so
$\mathfrak S^{(j)}$ is positive definite. Moreover,
\begin{equation}
    \left\|
    (\mathfrak S^{(j)})^{-1/2}-I
    \right\|_2
    \leq
    (1-\rho_j)^{-1/2}-1 .
    \label{eq:Appendix_D_inverse_bound}
\end{equation}

With the convention
$\mathfrak S_{\beta,\beta'}^{(j)}=\Braket{r_{\alpha,\beta}^{(j)}|r_{\alpha,\beta'}^{(j)}}$,
the L\"owdin-orthogonalized functions are written as
\begin{equation}
    s_{\alpha,\beta}^{(j)}
    =
    \sum_{\beta'=0}^{T_j-1}
    r_{\alpha,\beta'}^{(j)}
    \left[
    (\mathfrak S^{(j)})^{-1/2}
    \right]_{\beta',\beta}.
    \label{eq:Appendix_D_lowdin}
\end{equation}
This is the standard symmetric-orthogonalization convention.

For a coefficient vector $c$, the norm of
$\sum_{\beta}c_\beta r_{\alpha,\beta}^{(j)}$ is controlled by
$\mathfrak S^{(j)}$:
\begin{equation}
    \left\|
    \sum_{\beta=0}^{T_j-1}
    c_\beta r_{\alpha,\beta}^{(j)}
    \right\|_2^2
    =
    c^\dagger \mathfrak S^{(j)} c
    \leq
    (1+\rho_j)\|c\|_2^2 .
\end{equation}
Applying this to the coefficient vector given by the $\beta$-th column of
$(\mathfrak S^{(j)})^{-1/2}-I$, we find
\begin{equation}
\begin{aligned}
    \left\|
    s_{\alpha,\beta}^{(j)}
    -
    r_{\alpha,\beta}^{(j)}
    \right\|_2
    &\leq
    \sqrt{1+\rho_j}
    \left[
    (1-\rho_j)^{-1/2}-1
    \right]                                      \\
    &=O(\rho_j).
\end{aligned}
\label{eq:Appendix_D_state_difference}
\end{equation}
Therefore the perturbation caused by L\"owdin orthogonalization is exponentially
small in $h$ for fixed $T_j$.

It remains to connect this estimate with the localization widths. 
The $x$-space width is controlled by the weighted norm $\|xf\|_2$, while the
$k$-space width is controlled, under the symmetric Fourier convention, by the
corresponding derivative norm in $x$ space. 
Since the functions $r_{\alpha,\beta}^{(j)}$ are finite in number, compactly
supported, and smooth, the same coefficient estimate that gives
Eq.~\eqref{eq:Appendix_D_state_difference} also implies
\begin{equation}
\begin{aligned}
    &\left\|
    x\left(s_{\alpha,\beta}^{(j)}
    -
    r_{\alpha,\beta}^{(j)}\right)
    \right\|_2
    +
    \left\|
    \partial_x\left(s_{\alpha,\beta}^{(j)}
    -
    r_{\alpha,\beta}^{(j)}\right)
    \right\|_2                                      \\
    &\qquad
    \leq
    C_j \rho_j ,
\end{aligned}
\label{eq:Appendix_D_weighted_difference}
\end{equation}
where $C_j$ may depend on the finite block size $T_j$ and grows at most
polynomially in $h$ for fixed $T_j$. Since $\rho_j$ is exponentially small in
$h$, the changes in both localization widths can be made arbitrarily small by
taking $h$ sufficiently large. In particular,
\begin{equation}
\begin{aligned}
    &\left|
    \sigma_x(s_{\alpha,\beta}^{(j)})
    -
    \sigma_x(r_{\alpha,\beta}^{(j)})
    \right|
    +
    \left|
    \sigma_k(s_{\alpha,\beta}^{(j)})
    -
    \sigma_k(r_{\alpha,\beta}^{(j)})
    \right|                                      \\
    &\qquad
    \leq
    C_j'\rho_j ,
\end{aligned}
\label{eq:Appendix_D_width_difference}
\end{equation}
for another constant $C_j'$.

Finally, $r_{\alpha,\beta}^{(j)}$ is obtained from the truncated Gaussian
$\varphi$ by spatial translation and momentum modulation. These operations shift
the centers in $x$ and $k$ but do not change the corresponding variances.
From Eq.~\eqref{equ:Bourgain_initial}, $\varphi$ approaches the Gaussian
minimum-uncertainty packet as $h$ increases. Therefore, for any
$\varepsilon>0$, one can choose $h$ sufficiently large such that
\begin{equation}
    \sigma_x(r_{\alpha,\beta}^{(j)})
    <
    \frac{1}{\sqrt{2}}+\varepsilon,
    \qquad
    \sigma_k(r_{\alpha,\beta}^{(j)})
    <
    \frac{1}{\sqrt{2}}+\varepsilon .
    \label{eq:Appendix_D_r_bound}
\end{equation}
Combining Eqs.~\eqref{eq:Appendix_D_state_difference},
\eqref{eq:Appendix_D_width_difference}, and
\eqref{eq:Appendix_D_r_bound}, we obtain
\begin{equation}
    \sigma_x(s_{\alpha,\beta}^{(j)})
    <
    \frac{1}{\sqrt{2}}+2\varepsilon,
    \qquad
    \sigma_k(s_{\alpha,\beta}^{(j)})
    <
    \frac{1}{\sqrt{2}}+2\varepsilon .
    \label{eq:Appendix_D_final_bound}
\end{equation}
This shows that L\"owdin orthogonalization preserves the near-Gaussian
localization of the finite Bourgain lattice functions.
\begin{acknowledgments}
We thank Chunxi Zhang for helpful discussions about the Bourgain basis.
\end{acknowledgments}

\makeatletter\def\pre@bibdata{}\makeatother
\bibliography{reference}
\end{document}